\documentclass[twoside]{IEEEtran}
\usepackage{graphicx}
\usepackage{wrapfig}
\usepackage{amsfonts}
\usepackage{amssymb,epic,eepic}
\usepackage{amsmath}
\usepackage{dcolumn}
\usepackage{bm}
\usepackage{bbm}
\usepackage{verbatim}
\usepackage{color}
\usepackage{xargs}
\usepackage{tikz}
\usetikzlibrary{arrows,arrows.meta,positioning,decorations.pathmorphing,decorations.markings,shapes,fadings}

\newcommand{\eps}{{\varepsilon}}        

\newcommand{\eins}{{\mathbbm{1}}}

\newtheorem{theorem}{Theorem}

\newtheorem{definition}{Definition}

\newtheorem{lemma}[theorem]{Lemma}

\newtheorem{remark}[theorem]{Remark}

\newcommand{\tr}{\mathrm{tr}}

\DeclareMathOperator{\conv}{conv}

\tikzstyle{vecArrow} = [thick, decoration={markings,mark=at position
   1 with {\arrow[semithick]{open triangle 60}}},
   double distance=1.4pt, shorten >= 5.5pt,
   preaction = {decorate},
   postaction = {draw,line width=1.4pt, white,shorten >= 4.5pt}]
\tikzstyle{innerWhite} = [semithick, white,line width=1.4pt, shorten >= 4.5pt]

\tikzstyle{darkVecArrow} = [thick, decoration={markings,mark=at position
   1 with {\arrow[semithick,color=gray,shorten >= 1.5pt]{Triangle[scale=1]}}},
   double distance=1.4pt, shorten >= 5.5pt,
   preaction = {decorate},
   postaction = {draw,line width=1.4pt, gray,shorten >= 4.5pt}]

\title{Entanglement-Enabled Communication
\thanks{This work was funded by the DFG (grant NO 1129/2-1). Janis N\"otzel is with the Emmy-Noether Arbeitsgruppe Theoretisches Quantensystemdesign at the Lehrstuhl f\"ur Theoretische Informationstechnik of Technische Universit\"at M\"unchen, Germany (e-mail: janis.noetzel@tum.de).\textcolor{white}{..........................................................................................}
\textcolor{white}{-----------------------------------------------------------------------------------------}
\copyright 2020 IEEE. Personal use of this material is permitted. Permission from IEEE must be obtained for all other uses, in any current or future media, including reprinting/republishing this material for advertising or promotional purposes, creating new collective works, for resale or redistribution to servers or lists, or reuse of any copyrighted component of this work in other works.}
}
\author{
    \IEEEauthorblockA{Janis N\"otzel}
}

\begin{document}
\maketitle
\begin{abstract}
We  introduce and analyse a multiple-access channel with two senders and one receiver, in the presence of i.i.d. noise coming from the environment. Partial side information about the environmental states allows the senders to modulate their signals accordingly. An adversarial jammer with its own access to information on environmental states and the modulation signals can jam a fraction of the transmissions.
Our results show that for many choices of the system parameters, entanglement shared between the two senders allows them to communicate at non-zero rates with the receiver, while for the same parameters the system forbids any communication without entanglement-assistance, even if the senders have access to common randomness (local correlations). A simplified model displaying a similar behaviour but with a compound channel instead of a jammer is outlined to introduce basic aspects of the modeling.

We complement these results by demonstrating that there even exist model parameters for which entanglement-assisted communication is no longer possible, but a hypothetical use of nonlocal no-signalling correlations between Alice and Bob could enable them to communicate to Charlie again.

\end{abstract}
\begin{section}{Introduction}
What new possibilities does quantum nonlocality offer us? This question, that was posed in the 1994 publication \cite{popescu-rohrlich-1994} on quantum nonlocality, has not lost any of its appeal.

It is well-known that quantum technology offers dramatic advantages in the areas of computing \cite{Feynman1982}, secret communication \cite{BennetBrassardBB84}, randomness generation \cite{colbeck-thesis} and metrology \cite{bollingerEtAl-optimalFrequencyMeasurementsWithMaximallyCorrelatedStates}. It is not so well-known how quantum communication can increase the capacity of existing communication systems. Among the known mechanisms for providing such increase is \emph{superdense coding}, which was originally invented in \cite{bennett-wiesner}. In more general terms, this effect was studied under the name of entanglement assisted capacity of a quantum channel. Under practically reasonable restrictions on the encoding in the non-assisted case, it was shown in \cite{bennett-shor-smolin-thapliyal-eaCap} and restated in a simplified form in \cite{holevo-eaCap} that the difference between assisted and non-assisted capacity of a quantum channel can grow arbitrarily large. In terms of the potential technological impact of shared entanglement in the physical- or link layer of a communication network, these early findings provide sufficient insight to motivate further theoretical and practical study, as was for example carried out recently in \cite{guha-zhuang-boulat-infinitefold}. These works fall into the category of entanglement-assisted classical communication over quantum channels (category A).

In many of the known implementations of quantum effects, \emph{quantum entanglement} \cite{EPR,Schroedinger1935-theCat!} has been identified as the crucial enabling property providing the quantum advantage.

Yet, not much can be found on the potential of quantum entanglement as a plug-in resource for an otherwise completely classical communication system. To distinguish this category of systems from the aforementioned, we will use the term ``entanglement-assisted communication over classical channels'' (category B) for it. In the category B are multiple-access systems where signals travelling from a number of senders to a receiver have to use classical communication techniques, but a possibility exists for the senders to share and store entanglement. From a technical perspective, such systems can be very different from systems in category A. For example when the distance between senders is small compared to the distance from senders to receiver and no quantum repeater structure is available to distribute entanglement over the larger distance a system in category B could be realised, but none from category A. Conversely, if senders and receivers are close to each other it might make sense to use a system from category A, since the more advanced technology can be deployed on a short enough distance.

Among the publications in category B are for example the early work \cite{cubitt-leung-matthews-winter-ZeroErrorEntanglement2010}, as well as the more and very recent publications \cite{quekShorQuantumEnhancements} and \cite{leditzky2019playing}. The authors of \cite{cubitt-leung-matthews-winter-ZeroErrorEntanglement2010} studied zero-error transmission and gave an explicit example of a classical channel that could be used to transmit $5$ messages without error, but at least $6$ when shared entanglement between sender and receiver was allowed. In \cite{plmkr-entanglement-enhanced} an improvement from $5/6$ to $\tfrac{1}{3}(2+\tfrac{1}{\sqrt{2}})$ when using entanglement was demonstrated both theoretically and experimentally for another channel model. In the work \cite{quekShorQuantumEnhancements}, several interference channels were studied using analytical and numerical methods. It was proven that by explicit example that there are channels that have a purely classical rate region which is strictly smaller than the rate region when the communicating parties are allowed to access shared entanglement. Inspired by this work, the authors of \cite{leditzky2019playing} considered the multiple-access channel with shared entanglement between the senders. They proved that for their model, the largest possible sum rate using classical communication was upper bounded by $3.13694$, while the strategy using shared entanglement achieved a sum rate of $2\log3\approx3.17$. Both in \cite{quekShorQuantumEnhancements} and \cite{leditzky2019playing}, discrete memoryless systems of a rather simplistic type are studied. When comparing the status of research in category A with that in category B one finds that a missing piece in the category B is an answer to the question ``is there a bound on the helpfulness of quantum entanglement for systems in category B?''. To provide an answer to this question, we give a construction of a communication system in category B. The construction of the system is, at the cost of simplicity, made exactly such that the strongest possible difference between entanglement-assisted and non-assisted information transmission capabilities of the communication system is displayed - and answers the question raised by us to the negative.

The early literature dealing with the impact of entanglement on aspects of human interaction mostly compared classical- and quantum correlations in the setting of nonlocal games, where two or more players sitting in closed labs receive queries and have to give answers without communicating. The entanglement is thought of as enclosed in a black box with one interface per player. Each player operates its interface by choosing classical in- and reading classical outputs.

The prototypical nonlocal game, named after Clauser, Horne, Shimony and Holt (CHSH) \cite{chsh} who developed the original idea of Bell \cite{bell-epr}, has two players, each with binary inputs and binary outputs: $x,y\in\{0,1\}$, $\alpha,\beta\in\{0,1\}$. The players may agree on a strategy before the game starts, including a shared random variable $\lambda$; however, when play starts, they are separated, each receives their input (Alice $x$, Bob $y$), and without consulting each other, they have to respond with outputs $\alpha$ (Alice) and $\beta$ (Bob). Alice and Bob win the game if
\begin{equation}
  \label{eq:chsh}
  x y = \alpha \oplus \beta,
\end{equation}
otherwise they lose. The situation is depicted in Figure \ref{fig:chshGame}. Assuming uniform distribution on the inputs $x$ and $y$, \cite{chsh,bell-epr} showed that the maximum winning probability for classically correlated players is $3/4$, corresponding to the easily-verified fact that if $\alpha=\alpha(x)$ and $\beta=\beta(y)$ are functions of $x$ and $y$ alone, respectively, then Eq. (\ref{eq:chsh}) can be satisfied in only 3 out of 4 cases.
Interestingly, and crucially, with a quantum strategy, where each player holds one of two quantum bits (qubits) that are prepared in a maximally entangled state, and by making suitable quantum measurements on their respective systems, they can win with probability $\cos^2\frac{\pi}{8} = \frac12\left(1+\frac{1}{\sqrt{2}}\right) \approx 0.85 >3/4$. There are generalisations of the CHSH game, with more input and outputs
and different winning predicates \cite{Heywood1983,stairs1983,Mermin1990-QuantumMysteriesRevisited}.

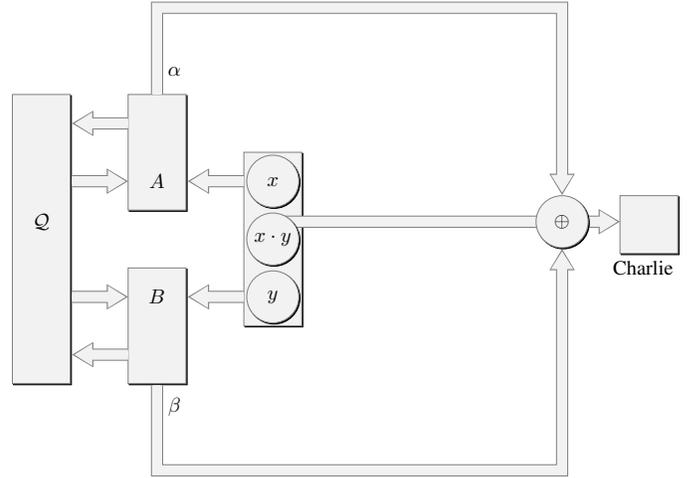
\begin{figure}
\centering
\definecolor{lightgray}{rgb}{.95,.95,.95}
\resizebox{.5\textwidth}{!}{
\begin{tikzpicture}[scale = 1]

\draw [fill=black,draw=black] (0.52,-2.52) rectangle (1.52,2.48);
\draw [fill=lightgray,draw=gray] (0.5,-2.5) rectangle (1.5,2.5);
\node (Resource) at (1,0.3) {$\mathcal Q$};

\draw [fill=black,draw=black] (2.52,2.48) rectangle (3.52,0.48);
\draw [fill=lightgray,draw=gray] (2.5,2.5) rectangle (3.5,0.5);
\node (A) at (3,1.0) {$A$};
\draw [fill=black,draw=black] (2.52,-2.52) rectangle (3.52,-0.52);
\draw [fill=lightgray,draw=gray] (2.5,-2.5) rectangle (3.5,-0.5);
\node (B) at (3,-1.0) {$B$};

\draw [fill=black,draw=black] (4.52,1.48) rectangle (5.52,-1.52);
\draw [fill=lightgray,draw=gray] (4.5,1.5) rectangle (5.5,-1.5);
\node (E) at (5,0.0) {$N_s$};

\draw [fill=black,draw=black] (11.02,0.73) rectangle (12.02,-0.27);
\draw [fill=lightgray,draw=gray] (11.0,0.75) rectangle (12.0,-0.25);
\node (C) at (11.4,-0.5) {Charlie};

\newlength{\myline}
\setlength{\myline}{1pt}
\newcommandx*{\triplearrow}[4][1=0, 2=1]{
  \draw[line width=1.5\myline,-{Triangle[scale=0.5]},draw = gray,double distance=3\myline] #4;
  \draw[line width=5\myline,-{Triangle[scale=0.5]},shorten <=#1\myline,shorten >=#2\myline,#3] #4;
}
\triplearrow{draw={-black!95}}{ (1.53,1) -- (2.5,1)};
\triplearrow{draw={-black!95}}{ (1.53,-1) to (2.5,-1)};
\triplearrow{draw={-black!95}}{ (2.5,2) to (1.53,2)};
\triplearrow{draw={-black!95}}{(2.5,-2) to (1.53,-2)};

\triplearrow{draw={-black!95}}{ (4.5,1) to (3.53,1)};
\triplearrow{draw={-black!95}}{ (4.5,-1) to (3.53,-1)};

\triplearrow{draw={-black!95}}{ (5.0,0.3) -- (11,0.3)};
\triplearrow{draw={-black!95}}{ (3,2.5) to (3,4) to (10,4) to (10,0.76)};
\triplearrow{draw={-black!95}}{ (3,-2.53) to (3,-4) to (10,-4) to (10,-0.17)};

\node at (5,0) {$x\cdot y$};
\draw [fill=black,draw=black] (5.02,0.98) circle (0.45);
\draw [fill=lightgray,draw=gray] (5,1) circle (0.45);
\node at (5,1) {$x$};
\draw [fill=black,draw=black] (5.02,-1.02) circle (0.45);
\draw [fill=lightgray,draw=gray] (5,-1.0) circle (0.45);
\node at (5,-1) {$y$};
\draw [fill=black,draw=black] (5.02,-0.02) circle (0.45);
\draw [fill=lightgray,draw=gray] (5,0) circle (0.45);
\node at (5,0) {$x\cdot y$};

\draw [fill=black,draw=black] (10.02,0.28) circle (0.45);
\draw [fill=lightgray,draw=gray] (10,0.3) circle (0.45);
\node at (10,0.3) {$\oplus$};

\node at (3.3,2.9) {$\alpha$};
\node at (3.3,-2.9) {$\beta$};
\end{tikzpicture}
}
\caption{\label{fig:chshGame}To enforce the technical aspect of the correlation, Alice and Bob are represented only via devices $A$ (for Alice) and $B$ (for Bob) which act on their behalf on input of the queries $x$ (for Alice) and $y$ (for Bob). The devices can interact (make measurements, receive measurement results) with a resource $\mathcal Q$. On receiving a measurement result, each device outputs a letter - $\alpha$ for Alice's device and $\beta$ for Bob's. Charlie acts as referee, receiving $x\cdot y\oplus\alpha\oplus\beta$. If his received letter is $0$, Alice and Bob won. If Charlie receives $1$, they lost.
}
\end{figure}

We show how to harness this advantage in a setting where Alice and Bob wish to communicate over a joint channel to a single receiver. The model is that of the multiple-access channel (MAC), which was introduced in the work of Shannon \cite{Shannon-two-way}, and solved by Ahlswede and Liao \cite{Ahlswede-Multiway,Liao-MAC-thesis}.

Our goal is to obtain the strongest possible separations between the entanglement-assisted and the purely classical communication system. It is known from the literature \cite{bennett-shor-smolin-thapliyal-eaCap} that shared entanglement cannot increase the capacity (in the sense of Shannon \cite{shannon-a-mathematical-theory-of-communication}) of a noisy channel. Earlier research has shown gains from distributed entanglement for classical communication using more complex channel models, but has not been able to prove the existence of an infinite gap. Therefore to demonstrate the existence of such a gap, we add a jammer who can influence the noisy channel in an ``arbitrary'' way without Alice and Bob knowing his action. Alice and Bob only get partial access to the channel state. The use of entanglement between two technical devices (called modulators) under the control of the sending parties is optional. We call this model the modulated arbitrarily varying multiple-access channel with (partial) environmental state information (MAVMACEI).

As is the case for the MAC, a noise process randomly generates different channel realizations. Here, this process is modeled by the random variable $X\cdot Y$ where both $X$ and $Y$ are sources of perfect random bits. Similar to the CHSH game, $X$ is made available to Alice's modulator and $Y$ to Bob's. Both modulators can modify the information transmitted by their respective owner depending on their received input. The channel states $X\cdot Y$ and the outputs $\alpha,\beta$ of the modulators are revealed to James, who then selects an additional input (state) to the channel. To limit his otherwise overwhelming capabilities, we subject James to a power constraint $\Lambda\in[0,1]$.

We show that there exist choices for the value of the power constraint, such that the rate region of our MAVMACEI consists of the (trivial) single point $\{(0,0)\}$ for purely classical coding but is nontrivial under entanglement-assisted coding. As the model consists of binary alphabets only, noise is modelled by random bit flips. A key observation then is that the jammer's state knowledge enables her to flip bits only at positions that were not in error already, thereby effectively increasing the probability of an error from e.g. $\tfrac{1}{4}$ to $\tfrac{1}{4}+\Lambda$. The MAVMACEI thus becomes useless for message transmission purpose as soon as $\Lambda\geq\tfrac{1}{4}$.

We forbid any communication between Alice and Bob. Instead, we allow for ``entanglement-modulated encoding''. When this method is used, a source of entangled quantum states is available to Alice and Bob. Based on their inputs $X$ or $Y$, the modulation units then perform a measurement on a shared quantum state. Their outputs $\alpha,\beta$ modify the sending parties input and can be read by James.

We show how such an entanglement-assisted coding scheme is able to reduce the effective initial noise to $\frac{2-\sqrt{2}}{4}$, such that the operating point at which the entanglement-assisted MAVMACEI cannot reliably transmit any messages any more is reached much later, when $\Lambda\geq\frac{1}{2\sqrt{2}}\approx0.35$.
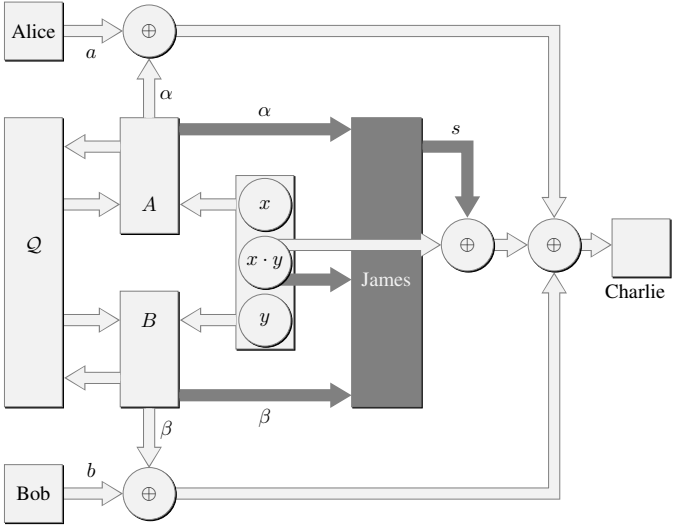
\begin{figure}
\centering
\definecolor{lightgray}{rgb}{.95,.95,.95}
\resizebox{.5\textwidth}{!}{
\begin{tikzpicture}[scale = 1]
\draw [fill=black,draw=black] (0.52,4.48) rectangle (1.52,3.48);
\draw [fill=lightgray,draw=gray] (0.5,4.5) rectangle (1.5,3.5);
\node (Alice) at (1,4.0) {Alice};

\draw [fill=black,draw=black] (0.52,-4.52) rectangle (1.52,-3.52);
\draw [fill=lightgray,draw=gray] (0.5,-4.5) rectangle (1.5,-3.5);
\node (Bob) at (1,-4.0) {Bob};

\draw [fill=black,draw=black] (0.52,-2.52) rectangle (1.52,2.48);
\draw [fill=lightgray,draw=gray] (0.5,-2.5) rectangle (1.5,2.5);
\node (Resource) at (1,0.3) {$\mathcal Q$};

\draw [fill=black,draw=black] (2.52,2.48) rectangle (3.52,0.48);
\draw [fill=lightgray,draw=gray] (2.5,2.5) rectangle (3.5,0.5);
\node (A) at (3,1.0) {$A$};
\draw [fill=black,draw=black] (2.52,-2.52) rectangle (3.52,-0.52);
\draw [fill=lightgray,draw=gray] (2.5,-2.5) rectangle (3.5,-0.5);
\node (B) at (3,-1.0) {$B$};

\draw [fill=black,draw=black] (4.52,1.48) rectangle (5.52,-1.52);
\draw [fill=lightgray,draw=gray] (4.5,1.5) rectangle (5.5,-1.5);
\node (E) at (5,0.0) {$N_s$};

\draw [fill=black,draw=black] (6.52,2.48) rectangle (7.72,-2.52);
\draw [fill=gray,draw=gray] (6.5,2.5) rectangle (7.7,-2.5);
\node (James) at (7.1,-0.3) {\textcolor{lightgray}{James}};

\draw [fill=black,draw=black] (11.02,0.73) rectangle (12.02,-0.27);
\draw [fill=lightgray,draw=gray] (11.0,0.75) rectangle (12.0,-0.25);
\node (C) at (11.4,-0.5) {Charlie};

\setlength{\myline}{1pt}
\newcommandx*{\triplearrow}[4][1=0, 2=1]{
  \draw[line width=1.5\myline,-{Triangle[scale=0.5]},draw = gray,double distance=3\myline] #4;
  \draw[line width=5\myline,-{Triangle[scale=0.5]},shorten <=#1\myline,shorten >=#2\myline,#3] #4;
}
\triplearrow{draw={-black!95}}{ (1.53,1) -- (2.5,1)};
\triplearrow{draw={-black!95}}{ (1.53,-1) to (2.5,-1)};
\triplearrow{draw={-black!95}}{ (2.5,2) to (1.53,2)};
\triplearrow{draw={-black!95}}{(2.5,-2) to (1.53,-2)};
\triplearrow{draw={-gray}}{ (3.53,2.3) to (6.5,2.3)};
\triplearrow{draw={-gray}}{ (3.53,-2.3) to (6.5,-2.3)};
\triplearrow{draw={-black!95}}{ (4.5,1) to (3.53,1)};
\triplearrow{draw={-black!95}}{ (4.5,-1) to (3.53,-1)};
\triplearrow{draw={-gray}}{ (5.0,-0.3) -- (6.5,-0.3)};
\triplearrow{draw={-black!95}}{ (10,0.3) -- (11,0.3)};
\triplearrow{draw={-black!95}}{ (8.7,0.3) -- (9.55,0.3)};
\triplearrow{draw={-black!95}}{ (5.0,0.3) -- (8.05,0.3)};
\triplearrow{draw={-black!95}}{ (3,-2.53) to (3,-3.56)};
\triplearrow{draw={-black!95}}{ (3,2.5) to (3,3.53)};
\triplearrow{draw={-black!95}}{ (1.53,4) to (2.55,4)};
\triplearrow{draw={-black!95}}{ (2.55,4) to (10,4) to (10,0.76)};
\triplearrow{draw={-black!95}}{ (1.53,-4) to (2.55,-4)};
\triplearrow{draw={-black!95}}{ (2.55,-4) to (10,-4) to (10,-0.17)};

\triplearrow{draw={-gray}}{ (7.73,2) to (8.5,2) to (8.5,0.77)};

\draw [fill=black,draw=black] (5.02,-0.02) circle (0.45);
\draw [fill=lightgray,draw=gray] (5,0) circle (0.45);
\node at (5,0) {$x\cdot y$};
\draw [fill=black,draw=black] (5.02,0.98) circle (0.45);
\draw [fill=lightgray,draw=gray] (5,1) circle (0.45);
\node at (5,1) {$x$};
\draw [fill=black,draw=black] (5.02,-1.02) circle (0.45);
\draw [fill=lightgray,draw=gray] (5,-1.0) circle (0.45);
\node at (5,-1) {$y$};
\draw [fill=black,draw=black] (3.02,3.98) circle (0.45);
\draw [fill=lightgray,draw=gray] (3,4.0) circle (0.45);
\node at (3,4) {$\oplus$};
\draw [fill=black,draw=black] (3.02,-4.02) circle (0.45);
\draw [fill=lightgray,draw=gray] (3,-4.0) circle (0.45);
\node at (3,-4) {$\oplus$};
\draw [fill=black,draw=black] (8.52,0.28) circle (0.45);
\draw [fill=lightgray,draw=gray] (8.5,0.3) circle (0.45);
\node at (8.5,0.3) {$\oplus$};
\draw [fill=black,draw=black] (10.02,0.28) circle (0.45);
\draw [fill=lightgray,draw=gray] (10,0.3) circle (0.45);
\node at (10,0.3) {$\oplus$};

\node at (3.3,2.9) {$\alpha$};
\node at (2,3.6) {$a$};
\node at (5,2.6) {$\alpha$};

\node at (3.3,-2.9) {$\beta$};
\node at (2,-3.6) {$b$};
\node at (5,-2.7) {$\beta$};

\node at (8.3,2.3) {$s$};
\end{tikzpicture}
}
\caption{\label{fig:channelmodel}The particular modulated arbitrarily varying multiple-access channel with environmental information that displays the strict separation between entanglement-assisted and non-assisted communication. Two bits $x$ and $y$ are generated uniformly at random by the environment. Alice knows $x$, Bob gets to know $y$. Depending on $x$ and $y$, the modulators $A$ and $B$ interact (e.g. by making measurements and observing measurement outcomes) with a resource $\mathcal Q$ that might consist of an entangled state. Alice transmits message $a$ and Bob transmits message $b$. The outputs of the modulators are added on top of the transmitted messages. A channel state is defined by $x\cdot y$. This channel state as well as the outputs $\alpha$ and $\beta$ of the modulators $A$ and $B$ are known to James. He uses $\alpha,\beta$ and $x\cdot y$ to come up with a bit $s$ that is added to the symbol $a\oplus b\oplus \alpha\oplus \beta$ originating from Alice, Bob and the modulators. Finally the channel state is added, so that Charlie receives $a\oplus b\oplus \alpha\oplus \beta\oplus s\oplus x\cdot y$. All additions are modulo $2$.
}
\end{figure}
\begin{remark}
    The communication model used in this work is motivated by the following observation: If we simply concatenate two classical discrete memoryless noisy channels, each having positive message transmission capacity, then the resulting channel can be expected to be more noisy, but - depending on the noise model - it might be hard to make it so noisy that it has a message transmission capacity of zero. An example are two binary symmetric channels with bit flip probabilities $p_1=p_2=\frac{1}{4}$. If they are concatenated, the resulting channel has a bit flip probability of $\tfrac{3}{8}\neq\tfrac{1}{2}$. Therefore, the concatenated channel still has a positive capacity.

    The goal of this work is to use entanglement solely as a plugin for coordination of communicating parties, and then to test the limits of this approach in terms of the systems capacity. The approach taken is to utilize, on the ``quantum side'' the established and tested mechanisms of the CHSH game for a communication scenario.

    If we let $x\cdot y$ from the CHSH game dictate the behaviour of a communication system in the sense that $x\cdot y=1$ triggers a bit-flip when Alice and Bob try to communicate to Charlie, then from analysis of the CHSH game we know that shared entanglement between Alice and Bob will increase their probability of jointly guessing $x\cdot y$. If we let them add the bits $\alpha$ and $\beta$ produced in response to $x$ and $y$ on top of $x\cdot y$ (compare Figure \ref{fig:chshGame}) then they will with probability $\approx0.85$ counter the bit flip triggered by $x\cdot y$. In addition, we can then let them add signals allowing them to transmit messages on top of $\alpha$ and $\beta$, and thus they will effectively transmit over a channel that first adds their signals together ($\mod{2}$) and then applies a binary symmetric channel with flip probability $\approx0.15$. If they were not using entanglement, that bit flip probability would be equal to $0.25$.

    While this approach is sufficient to demonstrate a quantum advantage, it does not maximize the difference between the assisted and the non-assisted system in terms of capacity (which is a function of the bit flip probability).

    An approach that appears intuitively useful is to search for methods of concatenating noisy channels that make two or more noise parameters behave \emph{additive}, and such that using entanglement reduces one of them below a threshold so that the sum becomes low enough to enable communication. As one can see from Theorem \ref{thm:jammer-knowledge-and-power-constraint}, the desired effect occurs for an arbitrarily varying jammer under a power constraint.

    An alternative to the way that we will lay out in full detail here is to create such a model by specifying a class of channels via parameters $\nu,\Lambda$ as follows: Let $BSC(p)$ denote a binary symmetric channel with bit flip probability $1-p$. Define
    \begin{align}
        Cl:=\{&\nu\cdot BSC(\omega_1)\cdot BSC(1)+\\
            &+(1-\nu)BSC(\omega_2)\cdot BSC(0):\nu,\omega_1,\omega_2\in[0,1]\}.\nonumber
    \end{align}
    Here, one can explicitly differentiate the channel actions in the first- and second stage of the model. The first stage sets the channel state, ($0$ or $1$) and the second acts depending on that state.

    Every channel within the class is a BSC, and every channel within the class is uniquely defined via its BSC parameter which is $\nu\omega_1+(1-\nu)(1-\omega_2)$. Within that class, one can then consider $\nu=\tfrac{3}{4}$, $\Lambda=\tfrac{2}{3}$ and study the subclass $Cl\upharpoonright(\nu=\tfrac{3}{4}\wedge\min\{\omega_1,\omega_2\}\geq\Lambda)$ of channels for which $\min\{\omega_1,\omega_2\}\geq\Lambda$. This is a convex set. Within this class is the channel where $\omega_1=\tfrac{2}{3}$ and $\omega_2=1$, and it has the BSC parameter $\tfrac{3}{4}\cdot\tfrac{2}{3} + \tfrac{1}{4}\cdot0=\tfrac{1}{2}$. Hence this BSC has capacity zero. Thus if one considers $Cl\upharpoonright(\nu=\tfrac{3}{4}\wedge\min\{\omega_1,\omega_2\}\geq\tfrac{2}{3})$ as a compound channel \cite{bbt-compound}, it has capacity zero. This extends to a chained model where one prepends a channel transforming two bits $(a,b)$ into their modulo two sum $a\oplus b$ and then attempts to transmit the sum over the compound channel. The analysis extends to all parameters $\nu\in[\tfrac{1}{4},\tfrac{3}{4}]$, by letting $\omega_2=\tfrac{2}{3}$ and $\omega_1=\tfrac{1}{2\nu} - \tfrac{1 - \nu}{3\nu}$ when $\nu\in[\tfrac{1}{4},\tfrac{1}{2}]$ and $\omega_1=\tfrac{2}{3}$, $\omega_2=\tfrac{1}{2(1-\nu)} - \tfrac{\nu}{3(1-\nu)}$ when $\nu\in[\tfrac{1}{2},\tfrac{3}{4}]$.

    Consider a setting where Alice and Bob, after receiving $x$ and $y$, encode bits $(a,b)$ (using potentially randomized encodings) into a channel $I$ that builds the modulo two sum of its binary inputs, and let this modulo two sum then be sent through the channel $BSC(\omega_{x\cdot y})\cdot BSC(x\cdot y)$. Lemma \ref{lem:restrictions-from-classical-strategies} shows that the best strategy that Alice and Bob can produce will effectively let them transmit $(a,b)$ over $Cl\upharpoonright(\nu=t\wedge \min\{\omega_1,\omega_2\}\geq\tfrac{2}{3})\circ I$ where $t\in[\tfrac{1}{4},\tfrac{3}{4}]$ is up for them to choose, and quantifies the residual noise after they used their information about $x$ and $y$ in the encoding. As outlined above, if the communicating parties cannot be sure which parameters $\omega_1,\omega_2$ out of the set of allowed $\omega_1,\omega_2$ are chosen, this multiple-access compound channel has the trivial rate region $\{(0,0)\}$.

    If one now considers entanglement as a way of reducing the noise $\nu$ in the first stage from $\nu=\tfrac{3}{4}$ to $\nu=\tfrac{1}{2}(1+\tfrac{1}{\sqrt{2}})$, as is suggested by the CHSH game, and keeps $\Lambda=\tfrac{2}{3}$, then one studies a second class of such channels where the function $(\omega_1,\omega_2)\to\nu\omega_1+(1-\nu)(1-\omega_2)$ is lower bounded by
    \begin{align}
        \tfrac{2}{3}\nu&=\tfrac{1}{3}(1+\tfrac{1}{\sqrt{2}})\\
        &=\tfrac{2}{6}+\tfrac{\sqrt{2}}{6}\\
        &>\tfrac{2}{6}+\tfrac{1}{6}\\
        &=\tfrac{1}{2}.
    \end{align}
    Since the set $Cl\upharpoonright(\nu=\tfrac{1}{2}(1+\tfrac{1}{\sqrt{2}})\wedge\min\{\omega_1,\omega_2\}\geq\Lambda)$ is convex, one can interchange $\min$ and $\max$ in the capacity formula $C=\max_p\min_{\omega_1,\omega_2\geq\Lambda}I(p;\nu\cdot BSC(\omega_1)+(1-\nu)BSC(1-\omega_2))$ proven in \cite{bbt-avc}. Thus the compound channel capacity of $Cl\upharpoonright(\nu=\tfrac{1}{2}(1+\tfrac{1}{\sqrt{2}})\wedge\min\{\omega_1,\omega_2\}\geq\Lambda)$ is that of a BSC with parameter $\tfrac{1}{3}(1+\tfrac{1}{\sqrt{2}})$, and therefore nonzero. Consequently, $Cl\upharpoonright(\nu=\tfrac{1}{2}(1+\tfrac{1}{\sqrt{2}})\wedge\min\{\omega_1,\omega_2\}\geq\Lambda)\circ I$ has a non-trivial rate region. Hence, we have motivated in detail the existence of at least one classical compound multiple-access channels which has a nontrivial rate region if the senders can use entanglement, but has the trivial rate region $\{(0,0)\}$ if they do not have access to entanglement.

    Building on this preliminary intuition, we will in the remainder of this work develop a slightly different channel model where instead of a compound channel we use an arbitrarily varying channel. We will thus prove in full detail the existence of communication systems in category B that have nonzero message transmission capacity only if augmented with quantum entanglement as an additional resource. In terms of the ratio between assisted- and non-assisted capacity, this is already the strongest possible difference that we could ever hope for, and thus the question is answered in the desired extremal way.
\end{remark}

To allow for a fair comparison, we study a third situation where Alice and Bob share ``local correlation'' instead of an entangled state.
In all cases, given the input $s=s(\alpha,\beta,x,y)$ of James the channel output is
\begin{align}
c = a\oplus b\oplus x\cdot y\oplus \alpha\oplus \beta\oplus s.
\end{align}
\begin{remark}
In this particular situation, James observes $\alpha$ and $\beta$ and the channel from Alice and Bob to Charlie is influenced by $x$ and $y$ only via $x\cdot y$ and via $\alpha$ and $\beta$. Thus it is not relevant whether, in addition to $\alpha$ and $\beta$, James observes $x$ and $y$ (so that $s=s(\alpha,\beta,x,y)$) or only $x\cdot y$ (in which case $s=s(\alpha,\beta,x\cdot y)$. When connection the communication model outlined here to the particular model of the arbitrarily varying channel with environmental information in the proof of Lemma \ref{lem:an-achievable-rate-region}.
\end{remark}
\end{section}
\begin{section}{Notation and Definitions\label{sec:notation-and-definitions}}
For two elements $x,y\in\{0,1\}$, $x\oplus y$ denotes addition modulo two. Given a finite alphabet $\mathbf X$, the set of probability distributions on it is $\mathcal P(\mathbf X)$. The corresponding state space for quantum systems on a finite dimensional Hilbert space $\mathcal H$ is denoted $\mathcal S(\mathcal H)$. For an element $x\in\mathbf X$ the symbol $\delta_x$ denotes an element of $\mathcal P(\mathbf X)$ with the property $\delta_x(x')=1$ if and only if $x=x'$. We'll also make use of the Kronecker delta in its form as a function $\mathbf X\times\mathbf X\to\{0,1\}$. It is related to the distributions $\{\delta_x\}_{x\in\mathbf X}$ via $\delta(x,x'):=\delta_x(x')$ for all $x,x'\in\mathbf X$. The symbol $\pi$ denotes the unique distribution with the property $\pi(x)=|\mathbf X|^{-1}$ for all $x\in\mathbf X$. To save space, we may occasionally write $p_i$ instead of $p(i)$. The convex hull of a set $X$ is $\conv(X)$. $\mathbf X\times\mathbf Y$ is the cartesian product of $\mathbf X$ with $\mathbf Y$. Composite quantum systems are modelled on tensor products $\mathcal H\otimes\mathcal H'$. The $n$-fold composition $\mathbf X\times\ldots\times\mathbf X$ is written $\mathbf X^n$. If $x^n,y^n\in\{0,1\}^n$ then $x^n\oplus y^n$ is defined component-wise as $(x^n\oplus y^n)_i:=x_i\oplus y_i$.

The scalar product of $x,y\in\mathbb C^d$ is denoted $\langle x,y\rangle$. A positive operator-valued measurement (POVM) on a Hilbert space consisting of $\mathbb C^d$ together with the standard scalar product is a collection $(M_i)_{i=1}^I$ of non-negative (meaning that $\langle x,M_ix\rangle\geq0$ for all $i=1,\ldots,I$ and $x\in\mathbb C^d$) matrices such that $\sum_{i=1}^IM_i=\mathbbm1$, where $\mathbbm1$ is the identity map on $\mathbb C^d$. For pure states, we write $\Psi=|\psi\rangle\langle\psi|$ where $\psi$ is any vector such that $\langle \psi,\Psi\psi\rangle=1$. The trace of a matrix $M$ is $tr(M)$.

A classical channel $W$ with input alphabet $\mathbf X$ and output alphabet $\mathbf Y$ is defined by a matrix $(w(y|x))_{x\in\mathbf X,y\in\mathbf Y}$ where $w(\cdot|x)\in\mathcal P(\mathbf Y)$ for all $x\in\mathbf X$. The set of all such channels is denoted $C(\mathbf X,\mathbf Y)$. Specific channels are: The identity on $\{0,1\}$, denoted $\mathbbm1$. The bit-flip on $\mathbbm F$ on $\{0,1\}$, acting as $\mathbbm F\delta_x=\delta_{x\oplus 1}$. The binary symmetric channel (BSC) with parameter $\nu\in[0,1]$ acts as
\begin{align}\label{def:bsc}
BSC(\nu)=\nu\mathbbm1+(1-\nu)\mathbbm F.
\end{align}
The binary adder channel $I\in C(\{0,1\}^2,\{0,1\})$ acts as
\begin{align}
I(\delta_{x,y})=\delta_{x\oplus y}.
\end{align}
We can incorporate external influence by setting
\begin{align}\label{def:J_0-and-J_1}
J_0:=\mathbbm1,\qquad J_1:=\mathbbm F
\end{align}
and defining an arbitrarily varying channel (AVC) as
\begin{align}
\mathcal J = (J_{s}\circ J_{y})_{s,y=0}^1.
\end{align}
In this model, the value of $s$ is dictated by a mechanism which is not under the control of the communicating parties. The exact nature of this control leads to further specification of the model. Typically the channel $\mathcal J$ itself is used in a memoryless fashion. However, the mechanism controlling $s$ might introduce memory into the model. For example, $s$ can be chosen by a so-called jammer who knows which code is used by senders and receiver and which message they intend to transmit, but not which code-word is used \cite{bbt-avc,boche-cai-cai-2018}. Alternatively, the jammer might know code, current code-word and current message \cite{ahlswede-wolfowitz-2} or only the code \cite{ahlswede-elimination,csiszar-narayan}. In another direction, recent research has started to introduce causality into the modelling \cite{dey-jaggi-langberg-sarwate-causal-avc}.

Throughout, $Wp$ is the output distribution of a channel $W$ upon input $p$ and $(W,p)$ the joint distribution of in- and output symbols:
\begin{align}
Wp(y)&:=\sum_xp(x)w(y|x),\\
(W,p)(y,x)&:=p(x)w(y|x).
\end{align}
The entropy of $p\in\mathcal P(\mathbf X)$ is $H(p):=-\sum_{x\in\mathbf X}p(x)\log(x)$, ($\log$ being calculated with base $2$ and using the convention $0\cdot\log(0)=0$). Whenever unambiguously possible, $p\in\mathcal P(\{0,1\})$ is identified with $p(1)$. In those cases we may write $h(p)$ or $h(p(1))$ to denote $H(p)$. The mutual information of $p\in\mathcal P(\mathbf X)$ and $W\in C(\mathbf X,\mathbf Y)$ is
\begin{align}
I(p;W):=H(p)+H(Wp)-H((W,p)).
\end{align}
For $\nu\in[0,1]$ we will typically use the abbreviation $\nu':=1-\nu$.

Given $n\in\mathbb N$ and a number $t$ satisfying $0\leq t\leq n$, the typical set $T^n_t\subset\{0,1\}^n$ is defined as $T^n_t:=\{x^n:N(1|x^n)=t\}$, where we set $N(x|x^n):=|\{i:x_i=x\}|$ for every $x\in\{0,1\}$ and $x^n\in\{0,1\}^n$. In cases where $n\in\mathbb N$ is clear from the context, the distribution $\pi_t\in\mathcal P(\{0,1\}^n)$ is defined as
\begin{align}
\pi_t(x^n):=\left\{
                \begin{array}{ll}
                    |T_t|^{-1},&x^n\in T_t\\
                    0,&\mathrm{else}
                \end{array}
            \right.\label{eqn:def-of-pi-t}
\end{align}
For $\delta\geq0$ and $p\in\mathcal P(\mathbf X)$ the $\delta$-typical set $T_{p,\delta}^n$ is defined as the set of all $x^n\in\mathbf X^n$ such that $|N(x|x^n)-n\cdot p|\leq n\delta$.

Let now $\mathbf A,\mathbf B,\mathbf C$ be the alphabets used by Alice, Bob and Charlie to en- and decode the messages.
\begin{definition}[Code\label{def:code}]
A code (for block length $n$) consists of message pairs $((u,v))_{u,v=1}^{U,V}$ and the corresponding code words $a^n_u,b^n_v\in \mathbf A^n,\mathbf B^n$ together with a collection $(D_{u,v})_{u,v=1}^{U,V}$ of decoding sets satisfying $D_{u,v}\subset \mathbf C^n$ for all $u,v\in[U],[V]$ and $D_{u,v}\cap D_{u',v'}=\emptyset$ whenever $(u,v)\neq(u',v')$.
\end{definition}
We will need local correlations, the study of which (in the context of attempts to understanding quantum entanglement) goes back to \cite{cirelson1980,popescu-rohrlich-1994}.
\begin{definition}\label{def:local-correlation}
A channel $Q\in C(\mathbf X\times\mathbf Y,\mathbf A\times\mathbf B)$ is called a ``local correlation'' if there is $p\in\mathcal P(\mathbf E)$ such that
\begin{align}
q(a,b|x,y)=\sum_{e}p(e)q_1(a|e,x)q_2(b|e,y).
\end{align}
It is called non-signalling if
\begin{align}
\forall\ a,x,y,y':\qquad \sum_bq(a,b|x,y)=\sum_bq(a,b|x,y')\\
\forall\ b,x,x',y:\qquad \sum_aq(a,b|x,y)=\sum_aq(a,b|x',y)
\end{align}
An example for a non-signalling correlation is the one presented in Lemma \ref{lem:epr-measurements}.
\end{definition}

\begin{definition}\label{def:AVMACEI}An arbitrarily varying multiple-access channel (AVMAC) with environmental information at the jammer (AVMACEI) consists of a channel $\mathcal W\in C(\mathbf A\times\mathbf B\times\mathbf S\times\mathbf Y,\mathbf C)$ and a $q\in\mathcal P(\mathbf Y)$. The probability for successful message transmission of a code $(D_{u,v})_{u,v=1}^{U,V}$ is
\begin{align*}
\min_S\frac{1}{UV}\sum_{u,v}\sum_{y^n,s^n}S(s^n|y^n)q^{\otimes n}(y^n)w^{\otimes n}(D_{u,v}|s^n,y^n,a^n_u,b^n_v).
\end{align*}
Minimization takes place over \emph{admissible} jamming strategies. $S$ is called admissible under a power constraint $\Lambda\geq0$ with constraint function $l$ ($l:\mathbf S\to\mathbb R_+$) if, for every $s^n$, $\sum_{i=1}^nl(s_i)> n\Lambda$ implies $S(s^n|y^n)=0$ for all $y^n$.
\end{definition}
\begin{remark}
    In the AVMACEI model, the environmental state is defined by $y\in\mathbf Y$, and the distribution of environmental states is $p\in\mathcal P(\mathbf Y)$. To put this model into perspective with the one used to show the extreme gap between the entanglement-assisted and non-assisted communication scenarios, it is depicted in Figure \ref{fig:genericAVMACEI}. To relate this model to the ordinary MAC, one can simply use $\mathbf S=\{1\}$, which leads to a simplification of the formula for the probability of successful message transmission to
    \begin{align*}
        \frac{1}{UV}&\sum_{u,v}\sum_{y^n}q^{\otimes n}(y^n)w^{\otimes n}(D_{u,v}|s^n,y^n,a^n_u,b^n_v)\\
        &=\frac{1}{UV}\sum_{u,v}\bar w^{\otimes n}(D_{u,v}|a^n_u,b^n_v)
    \end{align*}
    where $\bar w(c|a,b):=\sum_{y}p(y)w(c|1,y,a,b)$. This simplified formula is used to study message transmission over the MAC \cite{Ahlswede-Multiway,Liao-MAC-thesis}.
\end{remark}
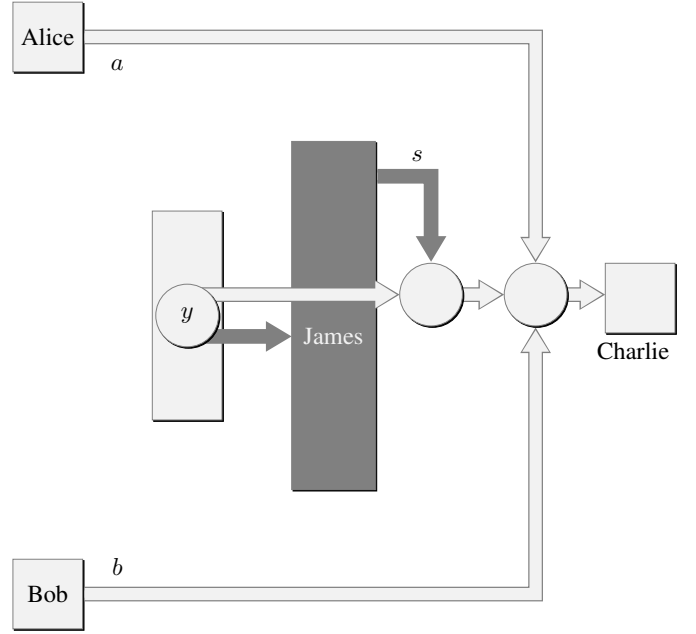
\begin{figure}
\centering
\definecolor{lightgray}{rgb}{.95,.95,.95}
\resizebox{.5\textwidth}{!}{
\begin{tikzpicture}[scale = 1]
\draw [fill=black,draw=black] (2.52,4.48) rectangle (3.52,3.48);
\draw [fill=lightgray,draw=gray] (2.5,4.5) rectangle (3.5,3.5);
\node (Alice) at (3,4.0) {Alice};

\draw [fill=black,draw=black] (2.52,-4.52) rectangle (3.52,-3.52);
\draw [fill=lightgray,draw=gray] (2.5,-4.5) rectangle (3.5,-3.5);
\node (Bob) at (3,-4.0) {Bob};

\draw [fill=black,draw=black] (4.52,1.48) rectangle (5.52,-1.52);
\draw [fill=lightgray,draw=gray] (4.5,1.5) rectangle (5.5,-1.5);
\node (E) at (5,0.0) {$N_s$};

\draw [fill=black,draw=black] (6.52,2.48) rectangle (7.72,-2.52);
\draw [fill=gray,draw=gray] (6.5,2.5) rectangle (7.7,-2.5);
\node (James) at (7.1,-0.3) {\textcolor{lightgray}{James}};

\draw [fill=black,draw=black] (11.02,0.73) rectangle (12.02,-0.27);
\draw [fill=lightgray,draw=gray] (11.0,0.75) rectangle (12.0,-0.25);
\node (C) at (11.4,-0.5) {Charlie};

\setlength{\myline}{1pt}
\newcommandx*{\triplearrow}[4][1=0, 2=1]{
  \draw[line width=1.5\myline,-{Triangle[scale=0.5]},draw = gray,double distance=3\myline] #4;
  \draw[line width=5\myline,-{Triangle[scale=0.5]},shorten <=#1\myline,shorten >=#2\myline,#3] #4;
}

\triplearrow{draw={-gray}}{ (5.0,-0.3) -- (6.5,-0.3)};
\triplearrow{draw={-black!95}}{ (10,0.3) -- (11,0.3)};
\triplearrow{draw={-black!95}}{ (8.7,0.3) -- (9.55,0.3)};
\triplearrow{draw={-black!95}}{ (5.0,0.3) -- (8.05,0.3)};

\triplearrow{draw={-black!95}}{ (3.53,4) to (10,4) to (10,0.76)};
\triplearrow{draw={-black!95}}{ (3.53,-4) to (10,-4) to (10,-0.17)};

\triplearrow{draw={-gray}}{ (7.73,2) to (8.5,2) to (8.5,0.77)};

\draw [fill=black,draw=black] (5.02,-0.02) circle (0.45);
\draw [fill=lightgray,draw=gray] (5,0) circle (0.45);
\node at (5,0) {$y$};

\draw [fill=black,draw=black] (8.52,0.28) circle (0.45);
\draw [fill=lightgray,draw=gray] (8.5,0.3) circle (0.45);

\draw [fill=black,draw=black] (10.02,0.28) circle (0.45);
\draw [fill=lightgray,draw=gray] (10,0.3) circle (0.45);

\node at (4,3.6) {$a$};

\node at (4,-3.6) {$b$};

\node at (8.3,2.3) {$s$};

\end{tikzpicture}
}
\caption{\label{fig:genericAVMACEI}A generic AVMACEI. Alice and Bob only know the statistical behaviour of the environment, which is described by $p\in\mathcal P(\mathbf Y)$. In contrast, James knows the exact realizations $y$. Dark arrows depict this information flow towards James. James uses his advanced knowledge to optimize his jamming strategy.
}
\end{figure}

\begin{definition}\label{def:AVMACEI-reg} A pair $(R_A,R_B)$ of non-negative numbers is called achievable for the AVMACEI $(\mathcal W,p)$ under power constraint $\Lambda$ with constraint function $l$ if there is a sequence $(\mathcal C_n)_{n\in\mathbb N}$ of codes with success probability according to Definition \ref{def:AVMACEI} going to $1$ as $n\to\infty$ and
\begin{align}
\liminf_{n\to\infty}\tfrac{\log U_n}{n}\geq R_A,\\
\liminf_{n\to\infty}\tfrac{\log V_n}{n}\geq R_B.
\end{align}
The rate region $\mathcal R^l_\Lambda(\mathcal W,p)$ of the AVMACEI is defined as the convex closure of the set of achievable rate pairs.
\end{definition}
Restricting the AVMACEI to one sender yields the AVCEI:
\begin{definition}\label{def:AVCEI}An Arbitrarily Varying Channel with Environmental Information (AVCEI) is a pair $(\mathcal W,p)$ where $\mathcal W\in C(\mathbf A\times\mathbf S\times\mathbf Y,\mathbf C)$ and $p\in\mathcal P(\mathbf Y)$. Given a code (with $V=1$), its probability of success for message transmission over $(\mathcal W,p)$ is
\begin{align*}
\min_S\frac{1}{U}\sum_{u=1}^U\sum_{y^n,s^n}S(s^n|y^n)p^{\otimes n}(y^n)w^{\otimes n}(D_u|a^n_u,s^n,y^n).
\end{align*}
Like for the AVMACEI, minimization is over all $S$ satisfying $\sum_{i=1}^nl(s_i)>n\Lambda\Rightarrow S(s^n|y^n)=0$ where $\Lambda\geq0$ is any given power constraint and $l:\mathbf S\to\mathbb R_+$.
\end{definition}
\begin{definition}\label{def:AVCEI-cap}A rate $R\geq0$ is called achievable for the AVCEI $(\mathcal W,p)$ with constraint function $l$ and power constraint $\Lambda$ if there is a sequence of codes such that
\begin{align}
\liminf_{n\to\infty}\tfrac{\log V_n}{n}\geq R.
\end{align}
The capacity $C^l_\Lambda(\mathcal W,p)$ is the supremum over all achievable rates.
\end{definition}
A key ingredient to our proofs is \cite[Theorem 3]{csiszar-narayan}.
The definition of the AVMACEI can be extended to a model where Alice and Bob share a communication resource. This resource may or may not be of a quantum mechanical nature, and this flexibility allows us to explain the usefulness of entanglement as a plug-in communication resource within an otherwise completely classical communication system.
\begin{definition}\label{def:MAVMACEI} A modulated AVMACEI with partial state information at the senders (MAVMACEI) is a MAC
$\mathcal W\in C(\mathbf E_A\times\mathbf E_B\times\hat{\mathbf Y}\times\mathbf M_\alpha\times\mathbf M_\beta\times\mathbf A\times\mathbf B\times\mathbf S,\mathbf C)$ together with a distribution $p\in\mathcal P(\mathbf E_A\times\mathbf E_B\times\hat{\mathbf Y})$. Abbreviating $\hat{\mathbf Y}\times\mathbf M_\alpha\times\mathbf M_\beta$ as $\mathbf Y$ shows the relation to the AVMACEI. Given any $\mathcal Q\in C(\mathbf E_A\times\mathbf E_B,\mathbf M_\alpha\times\mathbf M_\beta)$, the associated AVMACEI $(\hat{\mathcal W},\hat q)$ is defined by the conditional probabilities $\hat w(c|a,b,y,s)$ given by
\begin{align*}
\sum_{e_A,e_B}w(c|e_A,e_B,y,a,b,s)p'(e_A,e_B|\hat y).
\end{align*}
Here, $p'(e_A,e_B|\hat y):=p(e_a,e_B,\hat y)/p'(\hat y)$ with $p'(\hat y):=\sum_{e_A,e_B}p(e_A,e_B, \hat y)$. Without loss of generality, $p'(\hat y)>0$ for all $\hat y\in\hat{\mathbf Y}$. The distribution $\hat q$ equals $q(\alpha,\beta|e_A,e_B)p'(\hat y)$ .
\end{definition}

\begin{definition}A code for a MAVMACEI $(\mathcal W,p)$ is a shared resource $\mathcal Q\in C(\mathbf E_A\times\mathbf E_B,\mathbf M_\alpha\times\mathbf M_\beta)$ plus a code for the AVMACEI $\hat{\mathcal W}$ as in Definition \ref{def:MAVMACEI}. The code is called deterministic if $\alpha$ is a function only of $e_A$ and $\beta$ a function only of $e_B$. It is called ``jointly random modulated'' if $\mathcal Q$ is a local correlation. It is called entanglement-modulated if $q(\alpha,\beta|e_A,e_B)=tr(\Psi M_{A,e_A,\alpha}\otimes M_{B,e_B,\beta})$ for local measurements $M_{A,e_A}$ and $M_{B,e_B}$ and a quantum state $\Psi$.
\end{definition}
\begin{definition}\label{def:rate-region-of-the-MAVMACEI}$(R_A,R_B)$ is achievable for the MAVMACEI $(\mathcal W,p)$ under power constraint $\Lambda$ and cost function $l$ with
\begin{enumerate}
\item deterministically modulated codes
\item jointly random modulated codes
\item entanglement-modulated codes
\end{enumerate}
if there is a sequence $(\mathcal C_n)_{n\in\mathbb N}$ of corresponding codes such that ($R_A,R_B)$ is achievable for the AVMACEI $(\hat{\mathcal W},\hat q)$ using the corresponding modulation according to Definition \ref{def:AVMACEI}. The letters $\mathcal R_d^l((\mathcal W,p),\Lambda)$, $\mathcal R_r^l((\mathcal W,p),\Lambda)$, $\mathcal R_e^l((\mathcal W,p),\Lambda)$ denote the rate regions for the cases $1)$, $2)$ and $3)$.
\end{definition}
\begin{remark}\label{rem:specification-of-our-model}
The specific MAVMACEI studied here is denoted $\mathcal N$ (see Figure \ref{fig:channelmodel}). Every alphabet is binary, and
\begin{align*}
&p(e_A,e_B,\hat y)=\pi(e_A)\pi(e_B)\delta(\hat y,e_A\cdot e_B),\\
&w(c|e_A,e_B,y,\alpha,\beta,a,b,s)=\\
&\qquad=\delta(c,0\oplus 0\oplus e_A\cdot e_B\oplus \alpha\oplus \beta\oplus a\oplus b\oplus s).
\end{align*}
Here, $e_A$ and $e_B$ only influence the action of the modulators, while $\hat y=e_A\cdot e_B$ defines the environmental influence on the channel. We will write $i$ and $j$ instead of $e_A$ and $e_B$ in what follows to streamline notation.
\end{remark}
We achieve our separation result by showing the existence of values $\Lambda\in[0,1]$ of the power constraint such that only $(0,0)$ is achievable for $\mathcal N$ under jointly random modulated coding, while for the same values of the power constraint $\Lambda$ there are entanglement-modulated codes achieving strictly more rate pairs. Our entanglement modulation scheme is as follows:
\begin{definition}\label{def:epr-modulation}
Einstein-Podolsky-Rosen (EPR) modulation uses the state $\Psi=|\psi\rangle\langle\psi|$ for $\psi=\frac{1}{\sqrt{2}}(e_0\otimes e_1+e_1\otimes e_0)$ and measurements defined by the unitary matrices
\begin{align}
U_\theta:=\left(\begin{array}{ll}\cos\theta&\sin\theta\\-\sin\theta&\cos\theta\end{array}\right),
\end{align}
angles $\theta_0:=0,\ \theta_1:=\tfrac{\pi}{4},\ \tau_0:=\pi/8,\ \tau_1:=-\tfrac{\pi}{8}$ and setting - for all $i,j,x,y\in\{0,1\}$ -
\begin{align*}
M_{A,x,i}:=U_{\theta_x}|e_i\rangle\langle e_i| U_{\theta_x}^{-1},\ M_{B,y,j}:=U_{\tau_y}|e_j\rangle\langle e_j| U_{\tau_y}^{-1},
\end{align*}
where $\{e_0,e_1\}$ is the standard basis of $\mathbb C^2$.
\end{definition}

\end{section}
\begin{section}{Results and Proof of Main Result\label{sec:results-and-proof}}
We use the cost function $l(s)=s$ on $\{0,1\}$, $\Lambda\in[0,1]$, the MAVMACEI $\mathcal N$ (Remark \ref{rem:specification-of-our-model}), the AVMACEI $(\mathcal L,\nu)=(\mathcal J\circ I,\nu)$, and the AVCEI $(\mathcal J,\omega)$. The distributions $\nu,\omega$ can in principle be arbitrary, but our results will of course intertwine them in a particular way. Our main result is:
\begin{theorem}{\label{thm:main-result}}
The following are true:
\begin{enumerate}
\item $\mathcal R_d(\mathcal{N},\Lambda)=\mathcal R_r(\mathcal{N},\Lambda)=\{(0,0)\}$ for $\Lambda\geq\tfrac{1}{4}$
\item $\Lambda<\tfrac{1}{4}$ $\Rightarrow$ $\mathcal R_d(\mathcal{N},\Lambda), \mathcal R_r(\mathcal{N},\Lambda)\neq\{(0,0)\}$.
\item $\mathcal R_e(\mathcal{N},\Lambda)\neq\{(0,0)\}$ if $\Lambda<\tfrac{\sqrt{2}}{4}$.
\end{enumerate}
\end{theorem}
To prove Theorem \ref{thm:main-result}, we need some auxiliary statements. The first one reduces the channel that Alice and Bob can create to a concatenation of the adder channel with a BSC:
\begin{lemma}\label{lem:restrictions-from-classical-strategies}For all $A,B\in C(\{0,1\},\{0,1\})$ the MAC $L\in C(\{0,1\}^2,\{0,1\})$ defined by
\begin{align}
L:=\sum_{x,y}\sum_{i,j}\tfrac{A(i|x)B(j|y)}{4}J_{x\cdot y\oplus i\oplus j}\circ I
\end{align}
has the property $L\in\conv(\{BSC(\nu)\circ I:\tfrac{1}{4}\leq\nu\leq\tfrac{3}{4}\})$. For every $\mathbf E$ and $p\in\mathcal P(\mathbf E)$ the MAC defined via
\begin{align}
L_p:=\sum_{x,y}\sum_{e,i,j}\tfrac{p(e)A(i|e,x)B(j|e,y)}{4}J_{x\cdot y\oplus i\oplus j}\circ I
\end{align}
satisfies $L_p\in\conv(\{BSC(\nu)\circ I:\tfrac{1}{4}\leq\nu\leq\tfrac{3}{4}\})$.
\end{lemma}
Lemma \ref{lem:restrictions-from-classical-strategies} suggests the jammer, given his knowledge of the states $x\cdot y\oplus i\oplus j$, could realize a number of bit flips such that the number of bit flips in any transmission of $n$ bits over the channel roughly equals $\tfrac{n}{n}$ and the bit flips are placed randomly. This strategy can be sufficient to prohibit any communication over the AVMACEI:
\begin{lemma}\label{lemma:empty-rate-region}
Let the power constraint on James satisfy $\Lambda\geq\tfrac{1}{4}$ and let the parameter $\nu$ modelling the noise coming from the environment satisfy $\tfrac{1}{4}\leq\nu\leq\tfrac{3}{4}$. Then $\mathcal R_r^l((\mathcal L,\nu),\Lambda)=\{(0,0)\}$.
\end{lemma}
Lemma \ref{lemma:empty-rate-region} together with Lemma \ref{lem:restrictions-from-classical-strategies} implies statement $1)$ of Theorem \ref{thm:main-result} as follows: Let the channel $\mathcal N$ be given, together with a modulation strategy in the form of a conditional distribution $A(\alpha|e,i)$ for Alice and one $B(\beta|e,j)$ for Bob. If the environmental state $(x,y)$ is chosen uniformly at random, then
\begin{align}
w&(c|a,b,s):=\\
&:=\sum_{x,y}\sum_{i,j,e}\tfrac{p(e)A(i|e,x)B(j|e,y)}{4}\delta(c,x\cdot y\oplus i\oplus j\oplus a\oplus b\oplus s)\nonumber\\
&=\sum_{x,y}\sum_{i,j,e}\tfrac{p(e)A(i|e,x)B(j|e,y)}{4}[J_{x\cdot y\oplus i\oplus j\oplus s}(\delta_{a\oplus b})](c)
\end{align}
is the probability of receiving $c$ at the receiver when Alice's input is $a$ and Bob's input is $c$, while James chooses $s$. By Lemma \ref{lem:restrictions-from-classical-strategies}, both $w(c|a,b,0)=[BSC(\nu)\circ I(\delta_{a,b})](c)$ and $w(c|a,b,1)=[BSC(1-\nu)\circ I(\delta_{a,b})](c)$, where $\nu\in[\tfrac{1}{4},\tfrac{3}{4}]$. Here, we used the property $BSC(\nu)\circ\mathbb F=\mathbb F\circ BSC(\nu)$ which holds for all $\nu\in[0,1]$, together with $w(c|a,b,1)=w(c\oplus1|a,b,0)$. From Lemma \ref{lemma:empty-rate-region} we can thus conclude that $\mathcal N$ has an empty rate region if $\Lambda\geq\tfrac{1}{4}$.

To prove statements $2)$ and $3)$ in Theorem \ref{thm:main-result} we will make use of the following result:
\begin{theorem}\label{thm:jammer-knowledge-and-power-constraint}For the AVCEI $\mathcal J$ and every $\omega\in\mathcal P(\{0,1\})$:
\begin{align*}
&C_\Lambda^l(\mathcal J,\omega)=1-\underset{|\tau|\leq\Lambda}{\max}\ \eins_{[0,1]}(\omega+\tau)\cdot h(\omega+\tau).
\end{align*}
The following cases are particularly relevant here:
\begin{align*}
&C_{\tfrac{1}{4}}^l(\mathcal J,\tfrac{1}{4})=0, \ \ \ C_{\tfrac{1}{4}}^l(\mathcal J,\tfrac{2-\sqrt{2}}{4})=1-h(\tfrac{1+\sqrt{2}}{4})>0.
\end{align*}
\end{theorem}
\begin{remark}
The equality $C_{\tfrac{1}{4}}^l(\mathcal J,\tfrac{2-\sqrt{2}}{4})=1-h(\tfrac{1+\sqrt{2}}{4})$ is established as follows:
\begin{align}
C_{\tfrac{1}{4}}^l(\mathcal J,\tfrac{2-\sqrt{2}}{4})
&=1-\underset{|\tau|\leq\tfrac{1}{4}}{\max}\ \eins_{[0,1]}(\omega+\tau)\cdot h(\tfrac{2-\sqrt{2}}{4}+\tau)\nonumber\\
&=1-h(\tfrac{2-\sqrt{2}}{4}+\frac{1}{4})\\
&=1-h(\tfrac{1}{2}-\tfrac{2-\sqrt{2}}{4}-\frac{1}{4})\\
&=1-h(\tfrac{1}{4}-\tfrac{2-\sqrt{2}}{4})\\
&=1-h(\tfrac{1+\sqrt{2}}{4}).
\end{align}
where the second equality follows since $\tau\to\eins_{[0,1]}(\tfrac{2-\sqrt{2}}{4}+\tau)\cdot h(\tfrac{2-\sqrt{2}}{4}+\tau)$ is maximized for those values $\tau$ for which $|\tfrac{2-\sqrt{2}}{4}+\tau-\tfrac{1}{2}|$ is minimized, which is equivalent to minimizing $|\tfrac{\sqrt{2}}{4}-\tau|$ under the constraint $|\tau|\leq\tfrac{1}{4}$. Since $\sqrt{2}>1$ the minimum is attained at $\tau=\tfrac{1}{4}$. The third equality is a consequence of the symmetry of $x\to h(x)$ around $x=\tfrac{1}{2}$.

We will use the core idea in the proof of Theorem \ref{thm:jammer-knowledge-and-power-constraint} to prove Theorem \ref{thm:main-result}: If James knows the output of $p$ as well as those of the modulators, then he will effectively see an AVMACEI where the channel is either $\mathbbm1$ or $\mathbbm F$ (depending on the environmental variable $y$ and the outputs $\alpha,\beta$ of the modulators) acting on the joint input $a\oplus b$ of Alice and Bob.
\end{remark}
Theorem \ref{thm:jammer-knowledge-and-power-constraint} lets us prove that certain AVMACEI's have nontrivial rate regions by letting one sender send $0$'s only. Then, the channel for the other party is an AVCEI. Theorem \ref{thm:jammer-knowledge-and-power-constraint} gives its capacity $C$, proving achievability of $(C,0)$. This lets us conclude:
\begin{lemma}\label{lem:an-achievable-rate-region} The set of all $(R_A,0)$ and $(0,R_B)$ such that
\begin{align*}
0\leq R_A\leq C^l_\Lambda(\mathcal J,\tfrac{1}{4}),\qquad 0\leq R_B\leq C^l_\Lambda(\mathcal J,\tfrac{1}{4})
\end{align*}
is contained in $\mathcal R_d(\mathcal N,\Lambda)$, and thus $\mathcal R_d(\mathcal N,\Lambda)$ is strictly larger than $\{(0,0)\}$ whenever $\Lambda<\tfrac{1}{4}$.
\end{lemma}
Lemma \ref{lem:an-achievable-rate-region} implies statement $2)$ of Theorem \ref{thm:main-result}. To quantify the impact of entanglement we need to specify a quantum state and measurements for the modulators.
\begin{lemma}\label{lem:epr-measurements}Let $\psi=\tfrac{1}{\sqrt{2}}(e_0\otimes e_0+e_1\otimes e_1)$ and $\Psi=|\psi\rangle\langle\psi|$. Let $M_A,M_B$ be the measurements from Definition \ref{def:epr-modulation}. Set
\begin{align}
q(\alpha,\beta|x,y):=\tr(M_{A,x,\alpha}\otimes M_{B,y,\beta}\Psi)
\end{align}
(for all $\alpha,\beta,x,y\in\{0,1\}$). Then it holds
\begin{align}
q(\cdot|0,0)&=(\tfrac{1}{4}+t,\tfrac{1}{4}-t,\tfrac{1}{4}-t,\tfrac{1}{4}+t)\\
q(\cdot|1,0)&=q(\cdot|0,1)=q(\cdot|0,0),\\
q(\cdot|1,1)&=(\tfrac{1}{4}-t,\tfrac{1}{4}+t,\tfrac{1}{4}+t,\tfrac{1}{4}-t)
\end{align}
where outputs are indexed lexicographically and $t=\tfrac{1}{4\sqrt{2}}$.
\end{lemma}
These measurements can be used to realize what we would like to call ``EPR modulated encoding''. A benefit of this scheme is described in the following Lemma.
\begin{lemma}\label{lem:epr-modulated-encoding}The set of all $(R_A,0)$ and $(0,R_B)$ such that
\begin{align*}
0\leq R_A\leq C^l_\Lambda(\mathcal J,\tfrac{2-\sqrt{2}}{4}),\qquad 0\leq R_B\leq C^l_\Lambda(\mathcal J,\tfrac{2-\sqrt{2}}{4})
\end{align*}
is a subset of $\mathcal R_e(\mathcal N,\Lambda)$.
\end{lemma}
Lemma \ref{lem:epr-modulated-encoding} and Theorem \ref{thm:jammer-knowledge-and-power-constraint} prove statement $3)$ in Theorem \ref{thm:main-result}.
\begin{remark}
It is straightforward to verify that the rate region becomes even larger by using a nonlocal correlation taking the form as in Lemma \ref{lem:epr-measurements}, but with $t=\tfrac{1}{4}$ instead of $\tfrac{1}{4\sqrt{2}}$ - as defined in \cite[Equation (7)]{popescu-rohrlich-1994}. A look at equation \eqref{eqn:BSC-indexed-by-t-value} in the proof of Lemma \ref{lem:epr-modulated-encoding} verifies that for $t=\tfrac{1}{4}$ Alice and Bob will effectively transmit over the channel
\begin{align}
BSC(0)\circ I = I.
\end{align}
Such a scheme is thus able to cancel any noise coming from the environment and transform the channel into the binary adder channel $I$. The achievable rate region then contains the set as described in Lemma \ref{lem:epr-modulated-encoding}, but with $C^l_\Lambda(\mathcal J,\tfrac{2-\sqrt{2}}{4})$ replaced by $C^l_\Lambda(\mathcal J,0)$.
\end{remark}
\end{section}
\begin{section}{Proofs of Auxiliary Statements\label{sec:proofs-of-lemmata}}
\begin{IEEEproof}[Proof of Lemma \ref{lem:restrictions-from-classical-strategies}]
Let $A,B\in C(\{0,1\},\{0,1\})$. We will use symmetry arguments plus explicit evaluation of a few strategies to prove Lemma \ref{lem:restrictions-from-classical-strategies}. First, we consider extreme channels only. Under this restriction, our potential choices for $A$ and $B$ are limited to the set $E:=\{\mathbbm1,\mathbbm F,\mathbf0,\mathbf1\}$ where $\mathbf0(\delta_x)=\delta_0$ and $\mathbf1(\delta_x)=\delta_1$ for all $x\in\{0,1\}$. The transition probabilities from an input $(a,b)$ by Alice and Bob to an output $c$ for Charlie read as
\begin{align}
\mathbbm P(c|a,b)&=\sum_{x,y}\delta(c,x\cdot y\oplus Ax\oplus By\oplus a\oplus b).
\end{align}
The symbol $Ax$ is an abbreviation for the symbol $t$ such that $A(t|x)=1$, likewise $By$ is an abbreviation for the symbol $t$ such that $B(t|y)=1$.

Let us first consider a fixed choice of $a$ and $b$ such that $a\oplus b=0$, and choose $c=0$ as well. We need to understand the distributions of the random variables $
f_{A,B}(X,Y)$ where
\begin{align}
f_{A,B}(x,y):=x\cdot y\oplus Ax\oplus By.
\end{align}
If $(X,Y)$ are uniformly distributed, the distribution of $f_{A,B}(X,Y)$ is contained in $\conv(\{r,s\})$ for $r=(\frac{1}{4},\frac{3}{4})$ and $s=(\frac{3}{4},\frac{1}{4})$ - independent from the particular choice of $A,B\in E$.

To see this, we use the symmetry of $\pi\otimes\pi$ under exchange of $x$ with $y$ implying the statement needs only be proven for all $4$ choices $(A,B)$ of the form $(A,A)$ plus half of the remaining choices $(A,B)$ (since the results for e.g. $(\eins,0)$ equals that for $(0,\eins)$). It is also evident that the distribution of $f_{\eins,\eins}$ equals that of $f_{\mathbbm{F},\mathbbm{F}}$, and likewise for $f_{\mathbf1,\mathbf1}$. Since $\delta(a,b\oplus 1)=\delta(a\oplus 1,b)$ and $\mathbbm Fy=y\oplus 1$ we can easily confirm that the result holds for all choices $(A,B)\in\{(\mathbf1,\mathbbm{1}),(\mathbf1,\mathbbm{F}),(\mathbf0,\mathbbm1),(\mathbf0,\mathbbm{F})\}$ (since $r=\mathbbm Fs$) if it holds for only one of them. The same reasoning applies to the set $\{(\mathbf1,\mathbf0),(\mathbf0,\mathbf0),(\mathbf1,\mathbf1)\}$ and $\{(\mathbbm1,\mathbbm1),(\mathbbm1,\mathbbm{F})\}$. We thus set out to prove our result for the choices $\{(\mathbbm1,\mathbbm1),(\mathbbm1,\mathbf0),(\mathbf0,\mathbf0)\}$:
\begin{align}
\mathbb P&(f_{\eins,\eins}(X,Y)=0)=\tfrac{1}{4}\sum_{x,y}\delta(0,xy\oplus x\oplus y)\\
&=\tfrac{1}{4}(\delta(0,0)+\delta(0,1)+\delta(0,1)+\delta(0,1\oplus 1\oplus 1))\\
&=\tfrac{1}{4}.
\end{align}
\begin{align}
\mathbb P&(f_{\eins,\mathbf0}(X,Y)=0)=\tfrac{1}{4}\sum_{x,y}\delta(0,xy\oplus x)\\
&=\tfrac{1}{4}(\delta(0,0)+\delta(0,0)+\delta(0,1)+\delta(0,1\oplus 1))\\
&=\tfrac{3}{4}.
\end{align}
\begin{align}
\mathbb P&(f_{\mathbf0,\mathbf0}(X,Y)=0)=\tfrac{1}{4}\sum_{x,y}\delta(0,xy)\\
&=\tfrac{1}{4}\left(\delta(0,0)+\delta(0,0)+\delta(0,0)+\delta(0,1)\right)\\
&=\tfrac{3}{4}.
\end{align}
This reasoning together with our calculation demonstrates that
\begin{align}
\delta_z\mapsto \sum_{x,y}\tfrac{1}{4}\delta(\cdot,x\cdot y\oplus Ax\oplus By\oplus z)
\end{align}
equals $=BSC(\nu)$ for some $\tfrac{1}{4}\leq\nu\leq\tfrac{3}{4}$, depending on the choice of $(A,B)\in E$. Since the joint input $z$ by Alice and Bob is realized as $z=I(a,b)=a\oplus b$, the claim is proven for extremal strategies $A$ and $B$.
Thus the convex set $X$ of all possible effective channels that Alice and Bob can generate satisfies
\begin{align}
X \subset \conv(\{BSC(\tfrac{1}{4})\circ I,BSC(\tfrac{3}{4})\circ I\}).
\end{align}
That in fact $X=\conv(\{BSC(\frac{1}{4})\circ I,BSC(\frac{3}{4})\circ I\})$ can be seen by choosing the input $\mathbf0$ for Bob. In that case, Alice's choice $\mathbf0$ generates the effective channel $BSC(\tfrac{3}{4})$ and if Alice chooses $\mathbf1$ she generates the channel $BSC(\tfrac{1}{4})$.
This proves the proposed first claim of Lemma \ref{lem:restrictions-from-classical-strategies}.

The second claim follows by noting that $X$ is a convex set, and any random choice, including jointly random choices as the ones defined in Definition \ref{def:local-correlation}, of elements taken from $X$, will again produce an element of $X$.
\end{IEEEproof}
\begin{IEEEproof}[Proof of Lemma \ref{lemma:empty-rate-region}]
According to Lemma \ref{lem:restrictions-from-classical-strategies} there is a $\nu\in[\tfrac{1}{4},\tfrac{3}{4}]$ such that $BSC(\nu)\circ I=(1-\nu) J_0\circ I + \nu J_1\circ I$ is the effective MAC that Alice and Bob need to transmit over. For every two bits $(a_i,b_i)$ they send, $I$ converts them to $a_i\oplus b_i$ which then serves as an input to $J_{s}$ where
\begin{align}\label{def:s'}
s=x\cdot y\oplus \alpha\oplus \beta.
\end{align}
The bits $x,y$ are chosen according to $\pi\otimes\pi$. The probability that when $x,y$ are detected by the modulators a corresponding modulation signal $(\alpha,\beta)$ is created is
\begin{align}
q(\alpha,\beta|x,y):=\sum_{e\in\mathbf E}p(e)A(\alpha|e,x)B(\beta|e,y),\label{eqn:classical-strategy-form}
\end{align}
where $\mathbf E$ is a finite alphabet and $p\in\mathcal P(\mathbf E)$ (see Definition \ref{def:local-correlation}).

According to Definition \ref{def:MAVMACEI} James knows $s$ in \eqref{def:s'}. Thus for every strategy $q$ of the form \eqref{eqn:classical-strategy-form}, Lemma \ref{lem:restrictions-from-classical-strategies} states there will be a resulting $\nu\in[\tfrac{1}{4},\tfrac{3}{4}]$ such that for the purpose of analyzing the capacity region of the resulting channel, the AVMACEI
\begin{align}\label{eqn:avcei-is-the-correct-model}
(\mathcal L,\nu)=(\{J_s\circ J_y\circ I\}_{s,y=0}^1,\nu)
\end{align}
is the correct model. In order to show that for all choices $\nu\in[\tfrac{1}{4},\tfrac{3}{4}]$ the capacity region of $(\mathcal J,\nu)$ equals $\{(0,0)\}$ we can apply the jammer's strategy as described in Definition \ref{def:jammer-strategy} and applied in the proof of Theorem \ref{thm:jammer-knowledge-and-power-constraint}. Without loss of generality, assume that $\nu\in[\tfrac{1}{4},\tfrac{1}{2}]$. For the strategy from Definition \ref{def:jammer-strategy}, Lemma \ref{lem:jammer-strategy} and Lemma \ref{lem:distribution-of-noise} show that for every $\eps>0$ the impact of the strategy on the transmission from Alice and Bob to Charlie is effectively described by a distribution $\tilde p\in\mathcal P(\{0,1\}^n)$ (see equation \eqref{def:tilde-p} in the appendix) on the channel states with the following property: There is a sequence $(p'_n)_{n\in\mathbb N}$ of distributions $p'_n\in\mathcal P(\{0,1\})$ that satisfies $\lim_{n\to\infty}p'_n(1)=\pi_\eps(1)$ for $\pi_\eps\in\mathcal P(\{0,1\})$ defined via $\pi_\eps(1):=\tfrac{1}{2}-\eps$ such that for every $n\in\mathbb N$ it holds
\begin{align}
\tilde p_n\leq\tfrac{1}{1-\gamma_n}{p'_n}^{\otimes n}
\end{align}
for $\gamma_n=\delta_n(\eps)=2^{-nc(\eps)}$ where $c(\eps)>0$ is a suitable constant. Intuitively speaking, this particular jamming strategy simulates an i.i.d. distribution of the channel states $s\oplus y$ where $y$ is defined by the environment and $s$ by James. In particular, this i.i.d. distribution of the channel states is close to the uniform distribution. Thus letting $\eps_n:=-p'_n(1)+\tfrac{1}{2}$ we get for every two input strings $a^n,b^n$ and output string $c^n$ for Charlie we have
\begin{align}
\sum_{y^n,s^n}&\nu^{\otimes n}(y^n)J_{s^n+y^n}(\delta_{a^n}\otimes\delta_{b^n})(c^n)S(s^n|y^n)\label{eqn:start-upper-bounding-channel}\\
&=\sum_{\alpha^n}\tilde p(\alpha^n)J_{\alpha^n}(\delta_{a^n}\otimes\delta_{b^n})(c^n)\\
&\leq\tfrac{1}{1-\gamma_n}\sum_{\alpha^n}\pi_{\eps_n}^{\otimes n}(\alpha^n)J_{\alpha^n}(\delta_{a^n}\otimes\delta_{b^n})(c^n)\\
&=\tfrac{1}{1-\gamma_n}BSC(\tfrac{1}{2}-\eps_n)^{\otimes n}(\delta_{a^n\oplus b^n})(c^n).\label{eqn:end-upper-bounding-channel}
\end{align}
The last equality follows from the definition \eqref{def:J_0-and-J_1} of $J_0$ and $J_1$ together with the decomposition $BSC(\nu)=\nu J_0+(1-\nu)J_1$ resulting from the definition of a BSC in \eqref{def:bsc}.
For every $\eps>0$, $\lim_{n\to\infty}\gamma_n=\lim_{n\to\infty}\delta_n(\eps)=0$ and $\lim_{n\to\infty}\eps_n=\eps$. Thus for Alice and Bob, to transmit over $\mathcal J$ is asymptotically effectively equal to transmitting over $BSC(\tfrac{1-\eps}{2})\circ I$. It holds - for all $\rho,\sigma\in[0,1]$ -
\begin{align}
BSC(\rho)\circ BSC(\sigma)\circ I&=BSC(\rho\sigma+\rho'\sigma')\circ I\\
&=I\circ BSC(\rho)\otimes BSC(\sigma).
\end{align}
Choosing $\rho=\sigma=\tfrac{1}{2}-\sqrt{\tfrac{\eps}{2}}$, using the data processing inequality (c.f. Lemma 3.1 in \cite{csiszar-koerner}) and the capacity formula for the MAC we get for every pair of achievable rates $(R_A,R_B)$
and every $\eps>0$ that
\begin{align}
R_A+R_B&\leq \max_{p_1,p_2}I(p_1\otimes p_2;BSC(\tfrac{1}{2}-\eps)\circ I)\\
&=\max_{p_1,p_2}I(p_1\otimes p_2;BSC(\tfrac{1}{2}+\eps)\circ I)\\
&= \max_{p_1,p_2}I(p_1\otimes p_2;BSC(\rho)\circ BSC(\rho)\circ I)\\
&= \max_{p_1,p_2}I(p_1\otimes p_2;I\circ BSC(\rho)\otimes BSC(\rho) )\\
&\leq \max_{p_1,p_2}I(p_1\otimes p_2; BSC(\rho)\otimes BSC(\rho) )\\
&= 2-2h(\tfrac{1}{2}-\sqrt{\tfrac{\eps}{2}}).\label{inequ:avcei-has-empty-rate-region}
\end{align}
For values $\nu\in[\tfrac{1}{2},\tfrac{3}{4}]$ the strategy from Definition \ref{def:jammer-strategy} applies as well, the only modification is that James randomly selects his input on those channel states $y_i$ that are equal to one. Thus the rate region of $(\mathcal J,\nu)$ consists of the single element $\{(0,0)\}$, if the environmental state has a probability $\nu(1)\in[\tfrac{1}{4},\tfrac{3}{4}]$ and $\Lambda\geq\tfrac{1}{4}$.
\end{IEEEproof}
\begin{IEEEproof}[Proof of Theorem \ref{thm:jammer-knowledge-and-power-constraint}]
Our goal is to give a lower bound on the capacity $C^l_\Lambda(\mathcal J,\omega)$ by using the results of \cite{csiszar-narayan} and let that lower bound match an upper bound derived by explicitly quantifying the impact of one particular, valid jamming strategy (compare Defintion \ref{def:jammer-strategy} in the appendix).

To recapitulate the preliminaries, the power constraint $\Lambda\in[0,1]$ and the BSC parameter $\omega\in[0,1]$ are given from the statement of the Theorem. Without loss of generality we may assume that $\omega\neq\tfrac{1}{2}$, because for $\omega=\tfrac{1}{2}$ communication from sending to receiving party is impossible already without any active jamming. According to our convention, $\omega':=1-\omega$ and $\Lambda':=1-\Lambda$.

The idea of the jamming strategy is as follows: under i.i.d. noise from the environment, the `original' channel states $y^n$ will most likely have a certain number $t$ of ones. Assume w.l.o.g. $t\leq \tfrac{n}{2}$. James will receive those states. He will then generate a string equal to zero wherever $y_i=1$, and with a random pattern of zeroes and ones wherever $y_i=0$. He will select this random pattern such that the sum of the original string $y^n$ and his string $s^n$ does on average look as if distributed i.i.d. according to a new distribution that is as close as possible to $\pi$, while obeying the power constraint $\sum_is_i\leq n\Lambda$.

Our goal is to give a lower bound on the capacity $C^l_\Lambda(\mathcal J,\omega)$ based on \cite{csiszar-narayan} and let it match an upper bound derived by using above strategy, which is described in Definition \ref{def:jammer-strategy} in the appendix.

Thus we first take a side-step and consider the AVC as treated in \cite{csiszar-narayan}. In this model, there is no knowledge about the environmental channel states present at the jammer. Thus after treating this established model, we return to the one treated here and connect the two.

According to \cite{csiszar-narayan} the calculation of the capacity of an AVC with a power constraint on the jammer requires the definition of a function $p\to\Lambda_0(p)$ (compare equation (2.13) in \cite{csiszar-narayan}). The details of \cite{csiszar-narayan} that are relevant to our work are explained in Section \ref{sec:connection-to-csiszar-narayan}. This definition requires us to first fix a function $l$ and then minimize over the entire set of symmetrizers (compare Definition \ref{def:symmetrizability} in Section \ref{sec:connection-to-csiszar-narayan}) of the AVC. We will study this function for the special class of binary AVCs that take the form
\begin{align}\label{def:avbsc-via-nu}
L_\nu=(BSC(\nu),BSC(1-\nu))
\end{align}
for some $\nu\in[0,1]$. The question arises, whether we can explicitly write down the set of all symmetrizers for such AVCs. It is clear that for each $\theta\in[0,1]$ the map $q(s|x):=BSC(\theta)(s|x)$ is a symmetrizer:
\begin{align}
\sum_{s=0}^1q(s|0)L_{\nu,s}(\delta_1)&=q(0|0)\left(\begin{array}{l}\nu'\\ \nu\end{array}\right)+q(1|0)\left(\begin{array}{l}\nu\\ \nu'\end{array}\right)\\
&=q(1|1)\left(\begin{array}{l}\nu'\\ \nu\end{array}\right)+q(0|1)\left(\begin{array}{l}\nu\\ \nu'\end{array}\right)\\
&=q(1|1)L_{\nu,1}(\delta_0)+q(0|1)L_{\nu,0}(\delta_0)\\
&=\sum_{s=0}^1q(s|1)L_{\nu,s}(\delta_0).
\end{align}
Moreover, since
\begin{align}
\left\{\left(\begin{array}{l}\nu'\\ \nu\end{array}\right),\left(\begin{array}{l}\nu\\ \nu'\end{array}\right)\right\}
\end{align}
is a linearly independent set whenever $\nu\neq\tfrac{1}{2}$, we know that $\{BSC(\theta):\theta\in[0,1]\}$ is the complete set of symmetrizers, for every AVC $L_\nu$, if $\nu\neq\tfrac{1}{2}$.

We can now proceed and calculate the function $\Lambda_0:\mathcal P(\{0,1\})\to\mathbb R$ for every AVC $L_\nu$. The specific form of our cost function $l$ which is defined via $l(s)=s$ leads to a function $\Lambda_0$ as follows:
\begin{align}
\Lambda_0(p)&:=\min_{0\leq\theta\leq1}\sum_{x=0,s}^1p(x)BSC(\theta)(s|x)l(s)\\
&=\min_{0\leq\theta\leq1}\sum_{x=0}^1p(x)BSC(\theta)(1|x)\\
&=\min_{0\leq\theta\leq1}p(0)\theta+p(1)(1-\theta)\\
&=\min\{p(0),p(1)\}.
\end{align}
We define the following sets so that capacity formulas (c.f. \cite[Theorem 3]{csiszar-narayan}) can be written more efficiently:
\begin{align}
P_\Lambda^S&:=\{q\in\mathcal P(\{0,1\}):q(1)\leq\Lambda\}\\
P_\Lambda^X&:=\{p:\Lambda_0(p)>\Lambda\}.
\end{align}
By \cite[part $1)$, Theorem 3]{csiszar-narayan} (using the symbol $C_\Lambda^l(L_\nu)$ to denote what is written as $C(1,\Lambda)$ there, and $g(x)=1$ for all $x\in\{0,1\}$ being our particular choice for the function $g$ in \cite{csiszar-narayan}) we get for the channels $L_\nu$ as defined in \eqref{def:avbsc-via-nu} the implication  $\max_p\min\{p_0,p_1\}>\Lambda$ implies
\begin{align}
C_\Lambda^l(L_\nu)=\max_{p\in P_\Lambda^X}\min_{q\in P_\Lambda^S}I(p;\textstyle\sum_iq(i)BSC(\nu(i)))>0.\label{eqn:abstract-entropic-formula-for-capacity}
\end{align}
By part $2)$ of the same theorem, we have
\begin{align}
\max_p\min\{p_0,p_1\}<\Lambda\Rightarrow C_\Lambda^l(L_\nu)=0.
\end{align}
Let $\Lambda<\tfrac{1}{2}$ so that $\tfrac{1}{2}\in P_{\Lambda'}^X$. For every $q\in\mathcal P(\{0,1\})$, the channel $\sum_iq_iBSC(\nu_i)$ ($\nu_0:=\nu$, $\nu_1:=1-\nu$) is a BSC. Further, $P_\Lambda^S$ and $P_\Lambda^X$ are convex. Therefore
\begin{align}
\max_{p\in P_{\Lambda}^X}\min_{q\in P_{\Lambda}^S}&I(p;\textstyle\sum_{i=0}^1q_iBSC(\nu_i))=\\
&=\min_{q\in P_{\Lambda}^S}\max_{p\in P_{\Lambda}^X}I(p;BSC(\textstyle\sum_iq_i\nu_i))\\
&=\min_{q\in P_{\Lambda}^S}I(\pi;BSC(\textstyle\sum_iq_i\nu_i))\\
&=\min_{q\in P_{\Lambda}^S}(1-h(\textstyle\sum_iq_i\nu_i))\\
&=1-\max_{0\leq\tau\leq\Lambda}h(\tau\nu_1+\tau'\nu_0).
\end{align}
and thus
\begin{align}
\max_p\min\{p_0,p_2\}>\Lambda\ \Rightarrow\ C_\Lambda^l(L_0)=1-\max_{\tau\leq\Lambda}h(\tau).
\end{align}
Since $\max_{p\in\mathcal P(\{0,1\})}\min\{p_0,p_1\}=\tfrac{1}{2}$, we get
\begin{align*}
\Lambda<\tfrac{1}{2}\ \Rightarrow\ C_\Lambda^l(L_0)=1-\max_{0\leq\tau\leq\Lambda}h(\tau).
\end{align*}
If $\Lambda\geq\tfrac{1}{2}$ our jamming strategy implies $C^l_\Lambda(L_0)=0$, thus for all $\Lambda\geq0$ we get
\begin{align}
C_\Lambda^l(L_0)=1-\max_{0\leq\tau\leq\Lambda}\eins_{[0,1]}(\Lambda)h(\tau).
\end{align}
Consequently, $\Lambda\to C_{\Lambda}^l(L_0)$ is continuous on $[0,1]$.

We can now return to the study of the AVCEI $(\mathcal J,\omega)$. Initially, the connection between $L_0$ and $(\mathcal J,\omega)$ is quite simple: We just let the channel states for $L_0$ be chosen at random according to the distribution $\omega$. Intuitively, random i.i.d. noise should be less harmful to communication than active jamming based on state knowledge of $L_0$. We quantify this intuition in Lemma \ref{lem:connection-to-csiszar-narayan}, which lets us state that for the AVCEI $L_0$ and every $\delta>0$
\begin{align}\label{ineq:delta-bound}
C_{\Lambda}^l(\mathcal J,\omega)\geq C_{\Lambda+\omega+\delta}^l(L_0).
\end{align}
Therefore given such a $\delta>0$:
\begin{align}\label{ineq:delta-bound-2}
C_{\Lambda}^l(\mathcal J,\omega)\geq 1-\max_{0\leq\tau\leq\Lambda+\omega+\delta}\eins_{[0,1]}(\Lambda+\omega+\delta)h(\tau).
\end{align}
Since $\delta>0$ can be arbitrarily small, continuity of the right hand side of \eqref{ineq:delta-bound-2} yields
\begin{align}\label{ineq:delta-bound-3}
C_{\Lambda}^l(\mathcal J,\omega)\geq 1-\max_{0\leq\tau\leq\Lambda+\omega}\eins_{[0,1]}(\Lambda+\omega)h(\tau).
\end{align}
For $\omega<\tfrac{1}{2}$, this can be reformulated as
\begin{align}\label{ineq:delta-bound-4}
C_{\Lambda}^l(\mathcal J,\omega)\geq 1-\max_{0\leq\tau\leq\Lambda}\eins_{[0,1]}(\omega+\tau)h(\tau+\omega).
\end{align}
To show that $C_{\Lambda}^l(\mathcal J,\omega)=C_{\Lambda+\omega}^l(L_0)$ and become able to conclude $C_\Lambda^l(\mathcal J,\omega)=1-\max_{\tau\leq\Lambda}\eins_{[0,1-\omega]}(\tau)h(\tau+\omega)$ we use the jamming strategy defined in Definition \ref{def:jammer-strategy}. We first pick an $\eps$ that, intuitively speaking, helps us approximating the \emph{distribution} $\omega$ with the jamming strategy obeying \emph{state constraint} $\omega$. The action of James plus the environmental noise will, together, effectively act as if i.i.d. additive noise with parameter
\begin{align}
\eta(\Lambda,\omega):=\min\{\tfrac{1}{2},\omega+\Lambda\}
\end{align}
was present. Then, Definition \ref{def:jammer-strategy} and Lemma \ref{lem:distribution-of-noise} yield sequences $(p'_n)_{n\in\mathbb N}$ and $(\eps_n)_{n\in\mathbb N}$ with the property $p'_n\leq\tfrac{1}{1-\gamma_n}p_n^{\otimes n}$ for large enough $n$, $p_n\in\mathcal P(\{0,1\})$ defined by $p_n(1)=\eta(\omega,\Lambda)-\eps_n$ and $\eps_n\to\eps$. Similar to the previous calculations \eqref{eqn:start-upper-bounding-channel} to \eqref{eqn:end-upper-bounding-channel}, the effective channel from Alice to Charlie is then upper bounded by
\begin{align}
\tfrac{1}{1-\gamma_n}BSC(\eta(\omega,\Lambda)-\eps_n)^{\otimes n}.
\end{align}
Since $\eps_n\to\eps$ and $\eps>0$ is arbitrary it follows $\forall\ \omega\in[0,\tfrac{1}{2})$:
\begin{align}
C^l_{\Lambda}(\mathcal J,\omega)\leq1-\max_{|\tau|\leq\Lambda}\eins_{[0,1]}(\tau+\omega)h(\tau+\omega).
\end{align}
The situation $\omega\in(\tfrac{1}{2},1]$ is dealt with by letting James apply his random disturbances to those indices $i$ where $y_i=1$ instead of those where $y_i=0$. Thus $\forall\ \omega\in[0,1]$:
\begin{align}
C^l_\Lambda(\mathcal J,\omega)\leq1-\max_{|\tau|\leq\Lambda}\eins_{[0,1]}(\tau+\omega)h(\tau+\omega).
\end{align}
In combination with inequality \eqref{ineq:delta-bound-4}, this proves Theorem \ref{thm:jammer-knowledge-and-power-constraint}.
\end{IEEEproof}
\begin{IEEEproof}[Proof of Lemma \ref{lem:epr-measurements}]
Let $d\in\mathbb N$ and $\psi\in\mathbb C^d\otimes\mathbb C^d$ be given by
\begin{align*}
\psi=\frac{1}{\sqrt{d}}\sum_{i=1}^d e_i\otimes e_i.
\end{align*}
Let $U:\mathbb C^d\to\mathbb C^d$ be any matrices. It holds
\begin{align*}
(U\otimes\eins)\psi
    &=\frac{1}{\sqrt{d}}\sum_{i=1}^dUe_i\otimes e_i\\
    &=\frac{1}{\sqrt{d}}\sum_{i,j=1}^dU_{j,i}e_j\otimes e_i\\
    &=\frac{1}{\sqrt{d}}\sum_{i,j=1}^d(U^\top)_{i,j}e_j\otimes e_i\\
    &=(\eins\otimes U^\top)\psi.
\end{align*}
If $U_\theta,U_\tau:\mathbb C^d\to\mathbb C^d$ are unitary matrices with only real entries, then even
\begin{align*}
(U_\theta\otimes U_\tau)\psi=(U_\theta\cdot U_\tau^{-1}\otimes\eins)\psi.
\end{align*}
Let $P_i=:|e_i\rangle\langle e_i|$ and $P_j:=|e_j\rangle\langle e_j|$. Then
\begin{align}
\langle &(U_\theta\otimes U_\tau)\psi,(P_i\otimes P_j)(U_\theta\otimes U_\tau)\psi\rangle=\label{eqn:measurement-probabilities-start}\\
&=\langle (U_\theta\cdot U_\tau^{-1}\otimes\eins)\psi,(P_i\otimes P_j)(U_\theta\cdot U_\tau^{-1}\otimes\eins)\psi\rangle\\
&=\tfrac{1}{d}\langle U_\theta\cdot U_\tau^{-1}e_j,P_i\cdot U_\theta\cdot U_\tau^{-1} e_j\rangle\\
&=\tfrac{1}{d}\langle P_i\cdot U_\theta\cdot U_\tau^{-1}e_j,P_i\cdot U_\theta\cdot U_\tau^{-1} e_j\rangle\\
&=\tfrac{1}{d}|(U_\theta\cdot U_\tau^{-1})_{i,j}|^2\label{eqn:measurement-probabilities-end}
\end{align}
We now define the desired measurements as follows. Alice chooses an angle $\theta\in\{\theta_0,\theta_1\}$, depending on whether her bit $x$ equals $0$ or $1$. Bob chooses an angle $\tau\in\{\tau_0,\tau_1\}$ depending on his input. Their measurements are then defined by
\begin{align}
M_{A,x,i}&:=U_{\theta_x}P_i U_{\theta_x}^{-1}\\
M_{B,y,j}&:=U_{\tau_y}P_j U_{\tau_y}^{-1}
\end{align}
for all $i,j,x,y\in\{0,1\}$. Thus following equations \eqref{eqn:measurement-probabilities-start} to \eqref{eqn:measurement-probabilities-end} their probability of measuring $(i,j)$ upon input $(x,y)$ is
\begin{align}
\mathbb P(i,j|x,y)&=\langle (U_{\theta_x}\otimes U_{\tau_y})\psi,(P_i\otimes P_j)(U_{\theta_x}\otimes U_{\tau_y})\psi\rangle\nonumber\\
&=\tfrac{1}{d}|(U_{\theta_x}\cdot U_{\tau_{y}}^{-1})_{i,j}|^2.
\end{align}
We specify $U_\theta$ further by setting $d=2$ and
\begin{align}
U_\theta:=\left(\begin{array}{ll}\cos(\theta)&\sin(\theta)\\-\sin(\theta)&\cos(\theta)\end{array}\right)
\end{align}
and picking specific angles:
\begin{align}
\theta_0&:=0,\qquad &\theta_1:=\tfrac{\pi}{4}\\
\tau_0&:=\tfrac{\pi}{8},\qquad &\tau_1:=-\tfrac{\pi}{8}.
\end{align}
Then the measurement probabilities evaluate to
\begin{align}
\mathbb P(\{00,11\}|0,0)&=\frac{1}{2}\left(|(U_{-\tfrac{\pi}{8}})_{0,0}|^2+|(U_{-\tfrac{\pi}{8}})_{1,1}|^2\right)\\
&=\cos\left(\tfrac{\pi}{8}\right)^2
\end{align}
\begin{align}
\mathbb P(\{00,11\}|0,1)&=\frac{1}{2}\left(|(U_{\tfrac{\pi}{8}})_{0,0}|^2+|(U_{\tfrac{\pi}{8}})_{1,1}|^2\right)\\
&=\cos\left(\tfrac{\pi}{8}\right)^2
\end{align}
\begin{align}
\mathbb P(\{00,11\}|1,0)&=\mathbb P(\{00,11\}|0,1)
\end{align}
\begin{align}
\mathbb P(\{01,10\}|1,1)&=\frac{1}{2}\left(|(U_{\tfrac{3\pi}{8}})_{0,1}|^2+|(U_{\tfrac{3\pi}{8}})_{1,0}|^2\right)\\
&=\sin\left(3\tfrac{\pi}{8}\right)^2.
\end{align}
It holds $\cos\left(\tfrac{\pi}{8}\right)^2=\tfrac{1}{2}(1+\tfrac{1}{\sqrt{2}})$ and since $\sin\left(3\tfrac{\pi}{8}\right)^2=\cos\left(\tfrac{\pi}{8}\right)^2$ we get the desired result.
\end{IEEEproof}
\begin{IEEEproof}[Proof of Lemma \ref{lem:epr-modulated-encoding}]
The channel from Alice and Bob to Charlie is - if James is inactive - described by
\begin{align}
BSC(\tfrac{1-4t}{2})\circ I.\label{eqn:BSC-indexed-by-t-value}
\end{align}
For the specific value $t=\tfrac{1}{4\sqrt{2}}$ we arrive at an effective MAC
\begin{align}
BSC(\tfrac{1}{2}(1-\tfrac{1}{\sqrt2}))\circ I = BSC(\tfrac{2-\sqrt2}{4})\circ I.
\end{align}
Assume Bob transmits zeroes only, such that $(\mathcal J,\tfrac{2-\sqrt2}{4})$ is the channel from Alice to Charlie. Apply Theorem \ref{thm:jammer-knowledge-and-power-constraint}, to derive positive communication rates for Alice. Then, switch the roles between Alice and Bob to get positive rates for Bob as well.
\end{IEEEproof}
\end{section}
\begin{section}{Conclusion\label{sec:conclusion}}
It was known that distributed access to quantum correlations can improve the data transmission rates of communication systems. Our results have shown that there even exist purely classical communication systems which can only transmit data if assisted by entanglement. We have thus provided additional motivation for quantum technology as a part of classical signal processing.
\end{section}
\begin{section}{Appendix\label{sec:appendix}}
\begin{subsection}{On the relation of this work to \cite{csiszar-narayan}\label{sec:connection-to-csiszar-narayan}}
In \cite{csiszar-narayan}, the model of an arbitrarily varying channel (AVC) with a power constraint on both sender and jammer is studied. The AVC is introduced as a family
\begin{align}\label{def:csiszar-narayan-avc}
    \{w(\cdot|\cdot,s),s\in\mathbf S\}
\end{align}
of channels with finite input alphabet $\mathbf X$, finite output alphabet $\mathbf Y$ and finite set $\mathbf S$ of possible states $s\in\mathbf S$. The number $w(y|x,s)$ is the probability that $y$ is received when $x$ is the channel input and $s$ the channel state. For length-$n$ sequences, the transmission probability is introduced as
\begin{align}
w^{\otimes n}(y^n|x^n,s^n)=\prod_{i=1}^nw(y_i|x_i,s_i).
\end{align}
A length-$n$ block code consists of a set $\{1,\ldots,M\}$ of messages, code-words $x_1^n,\ldots,x_M^n\in\mathbf X^n$, and a decoder $\Phi:\mathbf Y^n\to\{0,1,\ldots,M\}$. The probability of error for such a code when message $i$ is sent and $s^n$ is the actual channel state sequence is given by
\begin{align}
    e(i,s^n)=\sum_{y^n:\Phi(y^n)\neq i}w^{\otimes n}(y^n|x^n,s^n).
\end{align}
The average probability of error of the code if channel state $s^n$ is used is
\begin{align}
    \bar e(s^n)=\frac{1}{M}\sum_{i=1}^Me(i,s^n).
\end{align}
A number $R\geq0$ is called an achievable rate if for every $\eps>0$ and $\delta>0$ there is an $N\in\mathbb N$ such that for all $n\geq N$ there is a length-$n$ block code such that
\begin{align}
    \frac{1}{n}\log M\geq R-\delta,\qquad \max_{s^n\in\mathbf S^n}\bar e(s^n)<\eps.
\end{align}
The capacity $C$ is then the supremum over all achievable rates. For this model, it was proven that
\begin{theorem}[Ahlswede, \cite{ahlswede-elimination}]\label{thm:ahlswede-dichotomy}
Define $I(p):=\min_{q\in\mathcal(\mathbf S)}I(p;\sum_{s\in\mathbf S}q(s)w(\cdot|\cdot,s))$. Either $C=\max_{p\in\mathcal P(\mathbf X)}I(p)$ or $C=0$.
\end{theorem}
It was observed earlier in \cite{ericson} that every AVC that is symmetric under an exchange of $x$ and $s$ has zero capacity. This observation can be extended to the statement that all AVCs that are symmetrizable have capacity zero \cite{ericson}, where symmetrizability is defined as follows:
\begin{definition}[Symmetrizability]\label{def:symmetrizability}
If there exists a $U\in C(\mathbf X,\mathbf S)$ such that
\begin{align}\label{eqn:symmetrizability-equation}
\sum_{s\in\mathbf S}u(s|x')w(y|s,x)=\sum_{s\in\mathbf S}u(s|x)w(y|s,x')
\end{align}
for all $x,x'\in\mathbf X$, $y\in\mathbf Y$ then $W$ is called \emph{symmetrizable}.
\end{definition}
One main contribution of \cite{csiszar-narayan} was to establish symmetrizability as the criterion for deciding whether $C>0$, by proving that every non-symmetrizable AVC has $C>0$. They also observed that non-symmetrizability implies that $I(p)$ from Theorem \ref{thm:ahlswede-dichotomy} is non-zero for every $p\in\mathcal P(\mathbf X)$ with the property $p(x)>0$ for all $x\in\mathbf X$. They then went on defining an AVC with \emph{input} or \emph{state constraints} as an AVC \eqref{def:csiszar-narayan-avc} together with two functions $g:\mathbf X\to\mathbb R$ and $l:\mathbf S\to\mathbb R$ satisfying $\min_{x\in\mathbf X}g(x)=0$ and $\min_{s\in\mathbf S}l(s)=0$. They extended those functions to $\mathbf X^n$ and $\mathbf S^n$ for every $n\in\mathbb N$ by setting
\begin{align}
    g(x^n)&:=\frac{1}{n}\sum_{i=1}^ng(x_i),\\
    l(s^n)&:=\frac{1}{n}\sum_{i=1}^nl(s_i).
\end{align}
Assuming two such functions and an AVC as given, a rate is then defined a sachievable for the AVC under input constraint $\Gamma\geq0$ and state constraint $\Lambda\geq0$ if there exists a code where all code-words $x^n_m$ satisfy $g(x^n_m)\leq\Gamma$ with the property that for every $\eps>0$ and $\delta>0$ there is an $N\in\mathbb N$ such that for all $n\geq N$ it holds
\begin{align}
    \frac{1}{n}\log M\geq R-\delta,\qquad\max\{\bar e(s^n):l(s^n)\leq\Lambda\}\leq\eps.
\end{align}
The capacity of the AVC under state constraint $\Lambda$ and input constraint $\Gamma$ is denoted $C(\Gamma,\Lambda)$ and is defined as the supremum over all rates that are achievable under input constraint $\Gamma$ and state constraint $\Lambda$.

To answer the question whether $C(\Gamma,\Lambda)>0$ for a given pair $(\Gamma,\Lambda)$ a certain function turned out useful, namely
\begin{align}
    \Lambda_0(p):=\min_{U\in\mathrm{Symm}(W)}\sum_{x\in\mathbf X}\sum_{s\in\mathbf S}p(x)u(s|x)l(s),
\end{align}
where $\mathrm{Symm}(W)$ is defined to be the set of all $U\in C(\mathbf X,\mathbf S)$ that satisfy equation \eqref{eqn:symmetrizability-equation}. For non-symmetrizable AVCs, one sets $\Lambda_0(p):=\infty$. For symmetrizable AVCs, $\Lambda_0$ is continuous. To see the usefulness of this definition, we cite \cite[Theorem 3]{csiszar-narayan}:
\begin{theorem}\label{thm:csiszar-narayan-main-result}
    For any $\Gamma>0$, $\Lambda>0$:
    \begin{enumerate}
        \item If $\max\{\Lambda_0(p):g(p)\leq\Gamma\}<\Lambda$ then $C(\Gamma,\Lambda)=0$.
        \item If $\max\{\Lambda_0(p):g(p)\leq\Gamma\}>\Lambda$ then $C(\Gamma,\Lambda)=\max\{I_\Lambda(p):g(p)\leq\Gamma\wedge\Lambda_0(p)\geq\Lambda\}$ and $C(\Gamma,\Lambda)>0$.
    \end{enumerate}
    Here, $I_\Lambda(p)$ is defined as the minimum of the function $(q,p)\mapsto I(p;\sum_{s\in\mathbf S}q(s)w(\cdot|s,\cdot))$ over the set of $q\in\mathcal P(\mathbf S)$ for which $\sum_{s\in\mathbf S}q(s)l(s)\leq\Lambda$.
\end{theorem}
\begin{remark}
    In this document, we apply Theorem \ref{thm:csiszar-narayan-main-result} to a particular fully binary AVC with $\mathbf S=\mathbf X=\mathbf Y=\{0,1\}$. As the state constraint, we choose $l(s):=s$ for all $s\in\mathbf S$ with varying $\Gamma\in[0,1]$, $g(x):=0$ for all $x\in\mathbf X$ and $\Gamma=1$ (no input constraint).
\end{remark}
\end{subsection}

\begin{subsection}{Proofs of Auxiliary Statements}

\begin{lemma}[Connecting Capacities\label{lem:connection-to-csiszar-narayan}]
Consider the AVC $(\eins,\mathbbm F)$ with cost function $l:\{0,1\}\to\mathbb R_+$ defined by $l(s):=s$ and without constraints on the sender. For a power constraint $\kappa$ on the jammer, let $C^l_\kappa((\eins,\mathbbm F))$ denote the capacity of $(\eins,\mathbbm F)$ as defined in \cite{csiszar-narayan}. For every $\Lambda_1,\Lambda_2\geq0$, $ \delta>0$ it holds
\begin{align}\label{ineq:connection-to-csiszar-narayan}
C_{\Lambda_1}^l((\eins,\mathbbm F),\Lambda_2)\geq C_{\Lambda_1+\Lambda_2+\delta}^l((\eins,\mathbbm F)).
\end{align}
\end{lemma}
\begin{IEEEproof}
Let $\Lambda:=\Lambda_1+\Lambda_2+\delta$. If $\Lambda\geq1$ then the right hand side in \eqref{ineq:connection-to-csiszar-narayan} is zero and thus there is nothing to prove. Therefore assume $\Lambda\in[0,1]$ in what follows.

Let $R\geq0$ and $(\mathcal C_n)_{n\in\mathbb N}$ be any sequence of codes with rate $R$ for $(\eins,\mathbbm F)$ under power constraint $\Lambda$ and cost function $l(s)=s$. Decompose $\Lambda$ as $\Lambda=\Lambda_1+\Lambda_2+\delta$ where $\Lambda_1,\Lambda_2\geq0$ and $\delta>0$ can be arbitrarily small. Let the average success probability of the code given a jammer state $s^n$ be
\begin{align}
\overline{\mathrm{sp}}(s^n):=\sum_m\frac{w^{\otimes n}(D_m|x^n,s^n)}{M}.
\end{align}
For every $\eps>0$ there is $N_1\in\mathbb N$ such that for all $n\geq N_1$
\begin{align}
1-\eps&\leq\min_{l(s^n)\leq\Lambda}\overline{\mathrm{sp}}(s^n).
\end{align}
Let $p\in\mathcal P(\{0,1\})$ be defined by $p(1):=\Lambda_2$. Then (see e.g. \cite[Lemma 2.12]{csiszar-koerner}, \cite[Theorem 1.1]{kramer-now}) there is an $N_2\in\mathbb N$ such that for all $n\geq N_2$ the lower bound
\begin{align}\label{ineq:typicality-at-work}
p^{\otimes n}(T_{p,\delta}^n)\geq1-\delta
\end{align}
holds true. For each $u^n\in T_{p,\delta}^n$ and $s^n$ with the property $l(s^n)\leq\Lambda_1$ the equality $1\oplus1=0$ implies
\begin{align}\label{ineq:additive-error-estimate}
l(s^n\oplus u^n)\leq\Lambda_1+\delta+\Lambda_2=\Lambda.
\end{align}
Therefore, we get
\begin{align}
    \mathrm{sp}
        &=_1 \min_S\sum_{u^n}p^{\otimes n}(u^n)\sum_{s^n:l(s^n)\leq\Lambda_1}S(s^n|u^n)\overline{\mathrm{sp}}(u^n\oplus s^n)\\
        &=_2\sum_{u^n}p^{\otimes n}(u^n)\min_{l(s^n)\leq\Lambda_1}\overline{\mathrm{sp}}(u^n\oplus s^n)\\
        &=\sum_{u^n\in T_{p,\delta}^n}p^{\otimes n}(u^n)\min_{l(s^n)\leq\Lambda_1}\overline{\mathrm{sp}}(u^n\oplus s^n)\nonumber\\
        &\qquad+\sum_{u^n\in (T_{p,\delta}^n)^\complement}p^{\otimes n}(u^n)\min_{l(s^n)\leq\Lambda_1}\overline{\mathrm{sp}}(u^n\oplus s^n)\\
        &\geq\sum_{u^n\in T_{p,\delta}^n}p^{\otimes n}(u^n)\min_{l(s^n)\leq\Lambda_1}\overline{\mathrm{sp}}(u^n\oplus s^n)\\
        &\geq_3\sum_{u^n\in T_{p,\delta}^n}p^{\otimes n}(u^n)\min_{l(u^n\oplus s^n)\leq\Lambda}\overline{\mathrm{sp}}(u^n\oplus s^n)\\
        &\geq_4\sum_{u^n\in T_{p,\delta}^n}p^{\otimes n}(u^n)(1-\eps)\\
        &=p^{\otimes n}(T_{p,\delta}^n)(1-\eps)\\
        &\geq_5 (1-\delta)(1-\eps)\\
        &\geq 1-\eps-\delta.
\end{align}
Here, the equality $1$ is the formula for the error criterion of an AVCEI from Definition \ref{def:AVCEI}, and equality $2$ assumes the worst-case strategy is used per $u^n$. Inequality $3$ is a consequence of inequality \eqref{ineq:additive-error-estimate}. Inequality $4$ follows from the assumption that the code used here works well up to a power constraint $\Lambda$ for James. Inequality $5$, finally, is an application of \eqref{ineq:typicality-at-work}.
This latter inequality in combination with inequality \eqref{ineq:typicality-at-work} allows us to conclude that $R$ is achievable for the AVCEI $((\eins,\mathbbm F),\Lambda_2)$ under power constraint $\Lambda_1$. It follows
\begin{align}
C_{\Lambda_1}^l((\eins,\mathbbm F),\Lambda_2)\geq C_{\Lambda_1+\Lambda_2+\delta}^l((\eins,\mathbbm F))
\end{align}
for every $\delta>0$.
\end{IEEEproof}

Our strategy assumes James applies a certain level $\Lambda$ of disturbance to any incoming binary sequence that has a ratio of zeroes and ones within a predefined range. This strategy is defined in detail in Definition \ref{def:jammer-strategy}. To meet the requirement $\sum_{i=1}^ns_i\leq n\Lambda$ for every $s^n$, the strategy is non-causal. We assume causal versions can be formulated as well. For the particular version applied in this manuscript, James first checks whether a noise sequence $\alpha^n$ is typical in the sense that $N(1|\alpha^n)\in[t_1,t_2]$ where $t_1,t_2$ are suitably chosen. If that is the case, he randomly selects a number $k$ from the set $\{0,\ldots,K\}$ according to a distribution on this set that is to be defined later. He then wants to change $\alpha^n$ in a number $k':=\chi(K,t,k)$ of positions. Here, $\chi$ is a function that enables one to achieve three goals: First, to let $K$ be flexible enough such that enough mass of a desired i.i.d. target distribution of $s^n\oplus\alpha^n$ can be captured. Second, to generate the same output distribution independent from $t$. Third, to ensure the power constraint is respected. James selects $k'$ positions $i_1,\ldots,i_{k'}$ where $\alpha_i=0$. He prepares a bit string $s^n$ such that $s_{i_j}=1$ for $j=1,\ldots,k'$ and $s_i=0$, else. The resulting channel state is $s^n+\alpha^n$ and has a number $t+k'$ of ones. To mimic an i.i.d. distribution of $s^n+\alpha^n$, the distribution of the parameter $k$ has to be chosen appropriately.
\begin{definition}\label{def:jammer-strategy}
Let $\Lambda\in[0,1]$. Set $\Lambda_n:=\lfloor n\Lambda\rfloor$ and let $t_1\leq t_2$, $K$ be natural numbers such that $\Lambda_n - K - t_2 + t_1 \geq 0$ and $t_1+\Lambda_n\leq n$. For any $k\in\{0,1,\ldots,K\}$ we set
\begin{align}
\chi(K,t,k):=k - (t -t_1) + (\Lambda_n -K).
\end{align}
Let $t\in\mathbb N$ satisfy $t_1\leq t\leq t_2$ and $k\in\{0,\ldots,K\}$. Given any $\alpha^n\in T_t^n$ and an $s^{n-t}\in T^{n-t}_{\chi(K,t,k)}$ for some $k\in\{0,\ldots,K\}$ we define a new string $S^n\in\{0,1\}^n$ element-wise via
\begin{align}
S(s^{n-t},\alpha^n)_i:= \left\{
                        \begin{array}{ll}
                            0,&\mathrm{if}\ \alpha_i=1\\
                            s_{N(0|\alpha^i)}\oplus \alpha_i,&\mathrm{else}.
                        \end{array}
                    \right.
\end{align}
Given a selection $\lambda_1,\ldots,\lambda_K\geq0$ of real numbers satisfying $\sum_k\lambda_k=1$ the jammer strategy for $\alpha^n\in T_t^n$ with $t\in\{t_1,\ldots,t_2\}$ is defined as
\begin{align}
    \hat S(\cdot|\alpha^n):=\sum_{k=0}^K\sum_{s^{n-t}\in T^{n-t}_{\chi(K,t,k)}}\frac{\lambda_k}{|T^{n-t}_{\chi(K,t,k)}|}\delta_{S(s^{n-t},\alpha^n)}.
\end{align}
The complete strategy $S$ is to apply $\hat S$ whenever $\alpha^n\in \cup_{t=t_1}^{t_2} T_t$ and apply $\tilde S(\cdot|\alpha^n):=\delta_{(0,\ldots,0)}$, else.
\end{definition}
This strategy obeys the power constraint $\Lambda$: The number of ones in the sequence $s^n$ added to any given $\alpha^n$ is either equal to $\chi(K,t,k)$ for some $t\in\{t_1,\ldots,t_2\}$ and $k\in\{0,1\ldots,K\}$ or zero, by design. For $t\in\{t_1,\ldots,t_2\}$ we get
\begin{align}
    \chi(K,t,k)
        &\leq\chi(K,t_1,K)\\
        &=\Lambda_n\\
        &\leq n\Lambda.
\end{align}
In addition, the following Lemma holds:
\begin{lemma}\label{lem:jammer-strategy}
Let $n,\Lambda,t_1,t_2,K$ and $\lambda_1,\ldots,\lambda_K\geq0$ be as in Definition \ref{def:jammer-strategy}. Then it holds for every $t\in\mathbbm N$ with the property $t_1\leq t\leq t_2$ that
\begin{align}
\sum_{k=0}^K&\sum_{\alpha^n\in T_t^n}\sum_{s^{n-t}\in T_{\chi(K,t,k)}^{n-t}}\frac{\lambda_k}{|T_{\chi(K,t,k)}^{n-t}|\cdot|T^n_t|}\delta_{S(s^{n-t},\alpha^n)}\\
&=\sum_{k=0}^K\lambda_k\pi_{\chi(K,0,k)},
\end{align}
where $\pi_t$ is defined in \eqref{eqn:def-of-pi-t}.
\end{lemma}
\begin{IEEEproof}[Proof of Lemma \ref{lem:jammer-strategy}]
To recapitulate, the strategy at work here is for James to add, for a number of potential choices of $k\in\mathbb N$, a number $k$ of ones at random positions $i\in[n]$, but restrict himself to those where $\alpha_i=0$. Our interest is to quantify the distribution of the resulting sequence of identities and bit flip channels that Alice and Bob will need to transmit over. Such a sequence is in one-to-one correspondence with the corresponding state sequence $s^n\oplus \alpha^n$.

Consider $k$ as fixed for the moment. If a given $c^n\in\{0,1\}^n$ has the property $N(1|c^n)\neq t+k$ then James' strategy will assign probability zero to the event $\alpha^n\oplus s^n=c^n$. Thus, we may assume $N(1|c^n)=t+k$. We may then consider all partitions of $c^n$ into two parts where the first part has exactly $t$ elements which are all equal to one. Each such decomposition corresponds to one choice of $\alpha^n$ from $T_t$. Using this decomposition we can explicitly calculate the probability that $\alpha^n\oplus s^n=c^n$ given that $\alpha^n\in T_t$ as
\begin{align}
\sum_{\alpha^n\in T_t}&\frac{1}{|T_t|}\mathbb P(\alpha^n\oplus s^n=c^n)=\\
&=\frac{1}{|T_t|}\sum_{\alpha^n\in T_t}\sum_{s^n:S(s^n|\alpha^n)>0}\tfrac{1}{|T_k^{n-t}|}\delta(\alpha^n\oplus s^n,c^n)\label{eqn:james-strategy-start}\nonumber\\
&=\binom{n}{t}^{-1}\binom{t+k}{t}\binom{n-t}{k}^{-1}\\
&=\frac{t!(n-t)!(t+k)!k!(n-t-k)!}{n!t!k!(n-t)!}\\
&=\binom{n}{t+k}^{-1}\\
&=\frac{1}{|T_{t+k}|}.\label{james-strategy-end}
\end{align}
This particular strategy of James aims at producing the same distribution of $\alpha^n\oplus s^n$, which should ideally match a desired target i.i.d. distribution, for a wide range of potential types $t$ of $\alpha^n$ - preferably at least those which are typical for $\omega^{\otimes n}$, when the environmental channel states $\alpha^n$ are distributed according to $\omega^{\otimes n}$ as is the case in our application of the strategy to the AVCEI $(\mathcal J,\omega)$.

Given any type $t$ with the property $t_1\leq t\leq t_2$, James samples an integer $k$ from the set $[K]$ according to $\lambda_0,\ldots,\lambda_K$. He then applies a corresponding (random) number $\chi(K,t,k):=k-(t-t_1)+(\Lambda_n-K)$ of bit flips to the channel according to the definition of $S(s^n|\alpha^n)$. By definition of the procedure and the equations preceding \eqref{james-strategy-end}, given any $t$, this produces - independent from $t$ - the distribution
\begin{align}
p':=\sum_{k=0}^K\lambda_{k}\pi_{t+k-(t-t_1)+(\Lambda_n-K)}&=\sum_{k=0}^K\lambda_{k}\pi_{\chi(K,0,k)}
\end{align}
which is well defined as long as the constraints
\begin{align}
t_1+\Lambda_n\leq n,\ \ K\leq t_1+\Lambda_n,\ \ t_1+\Lambda_n\geq K +t_2
\end{align}
are satisfied. Further, $p'$ does not depend on $t$ in case the constraints are satisfied.
\end{IEEEproof}
To quantify the impact of the jamming strategy from Definition \ref{def:jammer-strategy}, we describe the effective distribution of the additive noise generated by an ``original'' i.i.d. distribution $\eta$ of noise plus application of the jamming strategy. Recall that we write $\eta$ both for the distribution $\eta\in\mathcal P(\{0,1\})$ and for the real number $\eta(1)\in[0,1]$.
\begin{lemma}[Distribution of Noise\label{lem:distribution-of-noise}]
Let $\eta\in(0,\tfrac{1}{2})$, $\Lambda\in(0,\tfrac{1}{2}-\eta]$ and $0<\eps<\tfrac{1}{2}\min\{\Lambda,\eta\}$.
For every $n\in\mathbb N$, set $t_1:=\lceil (\eta-\eps)n\rceil$ and $t_2:=\lfloor(\eta+\eps)n\rfloor$ and $K:=\lfloor n2\eps\rfloor$ and $\Lambda_n:=\lfloor n\Lambda\rfloor$ and
\begin{align}
p_n:=\frac{\Lambda_n+K+t_1}{n}.
\end{align}
The sequence $(\eps_n)_{n\in\mathbb N}$ defined by $\eps_n:=\Lambda+\eta-p_n$ is non-negative and converges to $\eps$.

Define for every $k\in\mathbb N$ a number $c_k=1$ if $k\in\{p_n-\lfloor\tfrac{K}{2}\rfloor,p_n+\lfloor\tfrac{K}{2}\rfloor\}$ and $c_k=0$ else. If $p_n-\lfloor\tfrac{K}{2}\rfloor\geq0$, we set for each $k\in\{0,\ldots,n\}$
\begin{align}\label{def:lambda_k}
\lambda_k:=\frac{c_k}{p_n^{\otimes n}(\sum_lc_lT_l^n)}p_n^{\otimes n}(T_k^n).
\end{align}
Then for every $t$ satisfying $t_1\leq t\leq t_2$ and every $c^n\in\{0,1\}^n$
\begin{align}
\sum_{\alpha^n\in T^n_t}\sum_{s^n}\frac{S(s^n|\alpha^n)}{|T^n_t|}\delta(c^n,\alpha^n\oplus s^n)=\tilde p(c^n)
\end{align}
for $\tilde p$ defined via
\begin{align}\label{def:tilde-p}
\tilde p:=\sum_kc_k\lambda_k\pi_k.
\end{align}
There is a positive number $c(\eps)$ and an $N\in\mathbb N$ such that for all $n\geq N$ we have
\begin{align}
\tilde p\leq\frac{1}{1-2^{-nc(\eps)}}p_n^{\otimes n}.
\end{align}
\end{lemma}
\begin{IEEEproof}
The properties of $(\eps_n)_{n\in\mathbb N}$ are obvious from its definition.

We assume for simplicity that $K$ is an even number, which can be achieved by choosing $\eps$ appropriately and does not stop us from choosing arbitrarily small $\eps>0$. The coefficients $\lambda_k$ in \eqref{def:lambda_k} are well-defined since $p_n+\lfloor \tfrac{K}{2}\rfloor\leq n$ and $p_n-\lfloor\tfrac{K}{2}\rfloor\geq0$ once $n\geq2\cdot(\Lambda+\eta)^{-1}$. We set $N=2\cdot(\Lambda+\eta)^{-1}$. Then, the particular structure of $\tilde p(c^n)$ follows from Lemma \ref{lem:jammer-strategy}.

By e.g. utilizing \cite[Lemma 2.3]{csiszar-koerner} or direct application of \cite[Theorem 1.1]{kramer-now} we get$\sum_kc_kp^{\otimes k}(T_k)\geq1-2^{-nc(\eps)}$ where $\lim_{\eps\to0}c(\eps)=0$ and $c(\eps)>0$ for all $\eps>0$. For every $c^n\in T^n_k$ with $k\in\{k_0-\lfloor \tfrac{K}{2}\rfloor,k_0+\lfloor \tfrac{K}{2}\rfloor\}$ we then have $\tilde p(c^n)=\xi(k)\cdot p_n^{\otimes n}(c^n)$ for an appropriately chosen $\xi(k)>0$. For all other $c^n$, $\tilde p(c^n)=0$.

\end{IEEEproof}

\end{subsection}
\end{section}
\ \\
\emph{Acknowledgement.}
J.N. thanks Andreas Winter for inspiring and supportive discussions, and the unknown reviewers of this manuscript for their constructive criticism.
\bibliographystyle{IEEEtran}
\bibliography{percolation}

\begin{IEEEbiography}[{\includegraphics[width=1in,height=1.25in,clip,keepaspectratio]{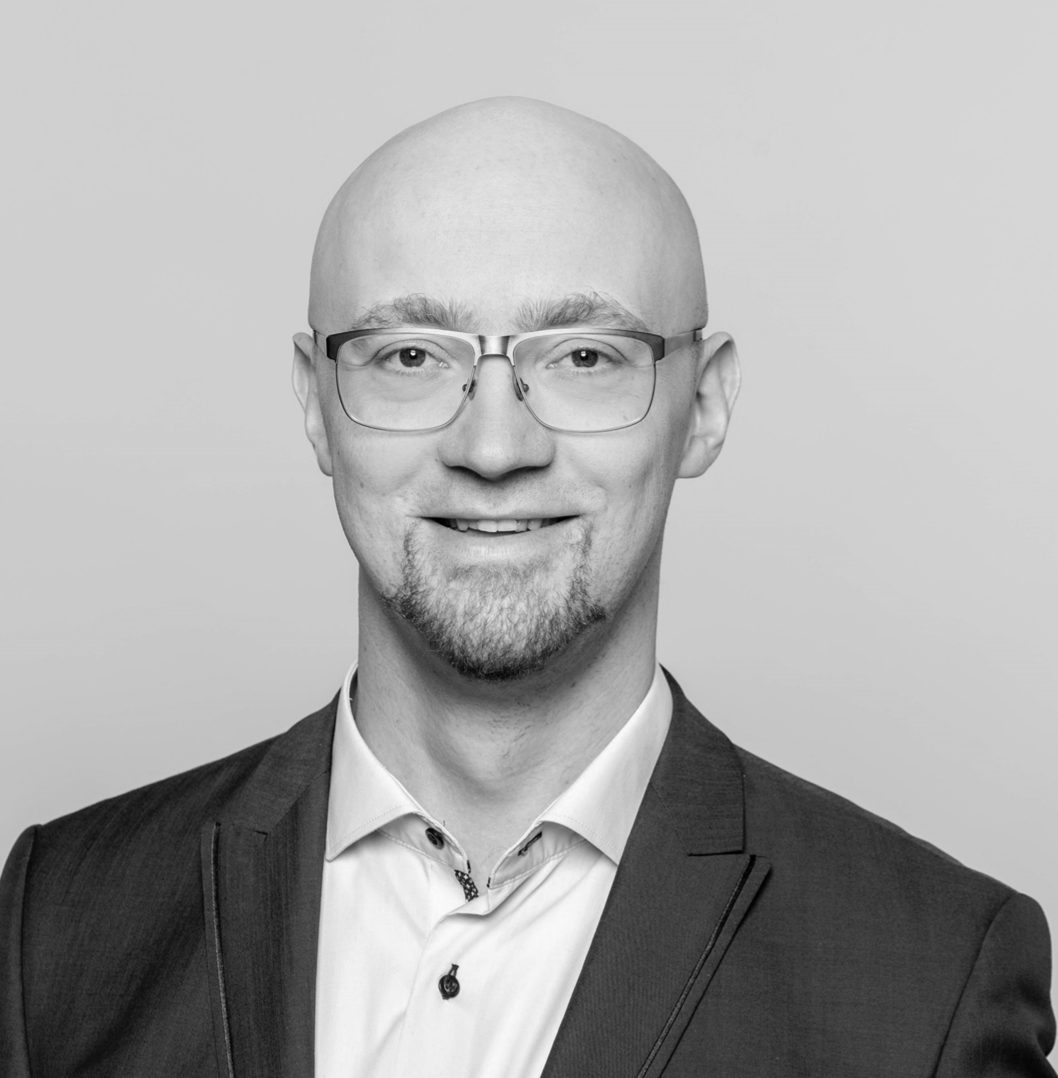}}]{Janis N\"otzel} received the Dipl. Phys. degree in physics from the Technische Universität Berlin, Germany, in 2007 and the PhD degree from Technische Universität München, Germany, in 2012.
From 2008 to 2010 he was a Research Assistant at the Technische Universität Berlin, Germany, and from 2011 until 2015 at Technische Universität München.
In 2015 and 2016 he was a DFG Research Fellow at Universitat Aut\`onoma de Barcelona, Spain.
From September 2016 to November 2018 he led a research transfer at the 5G Lab of the Technische Universität Dresden, Germany, resulting in the spin-off ZentiConnect.
In December 2018 he became an Emmy-Noether Research Group Leader at the Technische Universität München, Germany.
\end{IEEEbiography}
\end{document}